# Exploring the Future Metaverse: Research Models for User Experience, Business Readiness, and National Competitiveness

Amir Reza Asadi[1*] and Shiva Ghasemi[2]

[1*]School of Information Technology, University of Cincinnati, Cincinnati, 45221, Ohio, USA.

[2]Department of Computer Science, Virginia Tech, Blacksburg, 24060, Virginia, USA.

*Corresponding author(s). E-mail(s):
asadiaa@mail.uc.edu; Contributing authors:
Shivagh@vt.edu;

## Abstract

This systematic literature review paper explores perspectives on the ideal meta verse from user experience, business, and national levels, considering both academic and industry viewpoints. The study examines the metaverse as a sociotechnical imaginary, enabled collectively by virtual reality (VR), augmented reality (AR), and mixed reality (MR) technologies. Through a systematic litera ture review, n=144 records were included and by employing grounded theory for analysis of data, we developed three research models, which can guide researchers in examining the metaverse as a sociotechnical future of information technology. Designers can apply the metaverse user experience maturity model to develop more user-friendly services, while business strategists can use the metaverse business readiness model to assess their firms' current state and prepare for transformation. Additionally, policymakers and policy analysts can utilize the metaverse national competitiveness model to track their countries' competitive ness during this paradigm shift. The synthesis of the results also led to the development of practical assessment tools derived from these models that can guide researchers

# 1 Introduction

The term "Metaverse," originally coined by Neal Stephenson in his science fiction novel Snow Crash [1, 2], has transitioned from a futuristic concept to a key topic of interest in both industry and academia. The metaverse is not merely a 3D interactive internet [3], but a paradigm shift in social interactions that both imitates and con structs new sociocultural systems and information interactions [4–6]. The metaverse is a new reality and as the metaverse grows, it raises crucial questions about regu lation, privacy, and digital rights, underscoring the need for a multi-level analysis to inform comprehensive and effective policy-making.

The potential of the metaverse to reshape economies and societies necessitates a framework that can assess its implications across different scales. This complexity has led to the development of various models aimed at understanding the metaverse's intricacies. Jon Radoff's Metaverse value chain model [7], for instance, focuses on seven layers: Infrastructure, Human Interface, Decentralization, Spatial Computing, Creator Economy, Discovery, and Experience. Similarly, the Metaverse Maturity Index [8] evaluates Web3 readiness across capabilities, ambition, culture, technology, use cases, and governance. However, these models, while valuable, often address specific aspects of the metaverse in isolation.

There is a critical need for a comprehensive framework that integrates multiple perspectives and scales of analysis. The metaverse is developing at an unprecedented pace, with new technologies, applications, and use cases emerging constantly. A holistic framework is essential to track and understand these rapid changes across different domains.

Moreover, the interdisciplinary nature of the metaverse requires employing per spectives from both industry and academia. Current research often suffers from a disconnect between academic literature and industry insights. While academia pro vides theoretical frameworks, industry offers practical applications. Bridging this gap is crucial for a well-rounded understanding of the metaverse.

While there are many technological constraints in metaverse-enabling technolo gies, such as depth estimation challenges for seamless display of virtual content in augmented reality [9], motion sickness in virtual reality [10], and energy consump tion of portable devices [11, 12], focusing solely on these technical hurdles delivers an incomplete picture of metaverse landscape development. While solving these techno logical limitations is an essential and irreplaceable step toward realizing the metaverse, it is also crucial to consider it as a social transformation that will change people's interactions, disrupt business activities, and create new urban environments. This per spective requires a socio-political-economic-historical understanding [13] of how user experiences need to be transformed, how businesses need to be reorganized, and how government investment is required. In other words, we aim to explore the metaverse as a multifaceted phenomenon that encompasses not only technological advancements but also the broader societal, economic, and cultural implications of its adoption and integration into everyday life.

This study aims to address these challenges by developing an integrated framework that examines the metaverse from micro (individual user), meso (business), and macro (national) perspectives. We consider key factors such as user experience, business models, technological infrastructure, regulatory environments, and societal impact. By doing so, we seek to provide a comprehensive analysis that can inform both theoretical understanding and practical applications in the rapidly evolving metaverse landscape Our research is guided by three primary questions:

- RQ1: What are the primary elements affecting individual users' adoption and utilization of the metaverse?
- RQ2: What are the main factors driving the adoption and use of the metaverse within businesses?
- RQ3: What are the significant determinants impacting the adoption and utilization of the metaverse at the national level?

To answer this question, we seek to establish a framework that facilitates the under standing of these critical factors in a future-oriented manner, encompassing various scales from individual user experience to business impact and national significance.

In the following sections, we will report on the background of the topic, then outline the methodology used for the data collection, and analysis followed by a description of the models we developed based on the factors identified in the reviewed materials. Finally, we conclude by discussing potential directions for future research.

## 2 Background

This study is built on understanding metaverse as a sociotechnichal imaginary of infor mation technology [14, 15]. The study was guided by of Said et al. who defined information technology as " the study of solutions and needs that connect people, infor mation, and the technology of the time". We explore the metaverse as an integrated network of spatial realities and virtual worlds that constitutes a compelling alternative realm for human sociocultural interaction [17, 18]. The metaverse relies on early-stage and emerging technologies, including augmented reality (AR), virtual reality (VR), and mixed reality (MR), which are collectively referred to as extended reality (XR). XR, along with artificial intelligence (AI) and blockchain technology, are shaping a new digital reality where the boundaries between cyber and physical spaces are blurred [19]. This blending of realities is would be only achievable by ubiquitous comput ing, which provides the pervasive digital infrastructure and seamless human-computer interactions necessary for a cohesive, immersive metaverse experience[20].

In this social context, the convergence of community with information technol ogy will integrate communities, and social institutions [21] will create new needs for information that requires new solutions for connecting people.

Although XR technologies offer numerous applications, particularly in educa tion [22–26], the concept of the metaverse has yet to fully take shape. It remains a phenomenon we are likely to experience in the near future. This forward-looking aspect of our research presents distinct challenges, especially when it comes to identifying and evaluating key factors.

To address these challenges, we adopted methodological approaches from the development of e-readiness frameworks [27, 28] and maturity models [29, 30] in infor mation science studies. These approaches are particularly suitable for our they are designed to assess preparedness and potential for emerging technologies and paradigms.

# 3 Materials and Methods

A Systematic Literature Review (SLR) is a methodical approach for identifying, evaluating, and synthesizing existing primary research studies on a specific subject of interest. The research involves two stages of data collection following the PRISMA guidelines, an established methodology that aids in the identification and assessment of studies in various domains, including extended realities [31, 32]. For the analysis, the research followed grounded theory [33]. The application of PRISMA [34] was guided by an adaptation from Anwar and Azhar. The Figure 1 provides an overview of data collection.

## 3.1 Research Scope and Approach

Although this study aims to review the ideals about the metaverse applications, it does not cover the implementation aspects of enabler technologies for the metaverse such as XR and 5G. Since the concept of the metaverse impacts almost every field and situation, the preliminary analysis of articles about the Metaverse returned countless articles. To narrow the search, we applied Micro-Meso-Macro [35] as the analytical framework. This approach has previously been used for understanding ecosystems in HCI and information science studies.

## 3.2 Search Strategy

We collected data from academic databases by searching on Google Scholar and pro fessional networks by searching on Linkedin to incorporate both scholarly and industry viewpoints on the metaverse. Our data corpus included both white literature, such as academic articles, and grey literature that is not formally peer-reviewed, such as indus try reports, social media posts, and blogs [36]. This approach allowed us to capture the most current and diverse perspectives on the rapidly evolving metaverse phenomenon including both state of art and practice [37] by collecting non-academic articles too. The study collected first fifty results query searching on LinkedIn as previous study suggested including three first page of google search results [38].

By including white and grey literature, we were able to systematically explore both academic and industry perspectives on the metaverse. This dual focus allows us to capture theoretical insights from industry and triangulate them with academic findings, providing a more comprehensive view of the metaverse landscape. Through the qualitative analysis, we identified key variables that will serve as a foundation for future studies. These variables are essential in analyzing the ideals of the metaverse at the application, business, and national levels, thus contributing to a multidisciplinary understanding of this emerging phenomenon.

The selection of our search terms was driven by the necessity to thoroughly address our research questions. Our goal was to identify the keywords that would most effec tively support our investigation. The resulting search terms, as outlined in (see Table 1).

Table 1 Search Terms

| User Experience (Micro) | Business Readiness (Meso) | Governance (Macro) |
|---|---|---|
| Metaverse Experience Metaverse Design Metaverse UX Metaverse App in Metaverse | Metaverse Business Strategy Metaverse Maturity Metaverse Business Model | "Metaverse + Governance""Metaverse Governance" metaverse + "National Competitiveness" |

## 3.3 Screening and Selection

We applied the screening criteria outlined in Table 2. Through this systematic selec tion, we aimed to provide a comprehensive overview of factors that contribute to adaption of the metaverese.

## 3.4 Analysis

The analysis was adopts a grounded theory (GT) [39] approach which enables a rigorous and systematic approach to processing and analyzing existing research. This method involves open coding of existing data into relevant categories, supporting the development of new theories aimed at providing improved explanations and conceptual frameworks [40, 41].

The articles that satisfied the selection criteria were analyzed and coded in an iterative process. Coding is a crucial aspect of grounded theory, as it involves reviewing, interpreting, and identifying key phenomena within the collected data, then assigning codes to these phenomena to facilitate understanding [42]. The researchers followed the GLTR [33] for analysis by following the practice of Doyle and Brubaker. This approach involves open coding, axial coding, and selective coding. Open coding started by asking this question:"what is this corpus of literature interested in?" [43].

For instance, in the user experience domain, initial codes included "accessibility," "immersion," "interoperability," and "user empowerment." In the business domain, codes such as "digital twin " "metaverse,"resource and "product in virtual environ ments" emerged. For the national perspective, codes like "infrastructure" "regulation " and "investment" were identified. After the initial open coding, the researchers moved to axial coding. In this phase, relationships between the codes were identified and cat egories were formed. The final stage was selective coding, where core categories were identified and refined. These core categories formed the basis of the theoretical models developed in this study.

| Research Focus | Inclusion Criteria | Exclusion Criteria |
|---|---|---|
| User Experience Fac tors | • Studies on user experience in metaverse/VR environments • Research on accessibility and inclusion<br>• Literature on social and cul tural implications<br>• Articles on privacy, security, and identity | • Studies solely on gaming without metaverse context • Publications on VR/AR hardware without user expe rience focus |
| Business Factors | • Articles on business strate gies for metaverse adoption • Studies on metaverse eco nomics and virtual goods • Literature on digital twins in business contexts<br>• Articles on metaverse mar keting and branding | • Studies on traditional e-commerce without meta verse context<br>• Research on general digital transformation not specific to metaverse<br>• Studies on individual com pany competitiveness rather than national |
| Country Factors | • Articles on national policies and governance for meta verse<br>• Research on infrastructure for national metaverse adop tion<br>• Studies on metaverse regula tory frameworks<br>• Literature on economic impact of metaverse at national level | • Articles on general digital technologies without meta verse focus<br>• Articles that promote a political party campaign |

| 4. General Criteria | • Publications on meta verse technologies (AR/VR, blockchain)<br>• Papers on interoperability and standards<br>• Both academic and industry publications | • Articles not in English<br>• Opinion pieces without sub stantial research<br>• Duplicate studies<br>• Articles Published before 2015 |
| --- | --- | --- |

Table 2 Inclusion and Exclusion Criteria

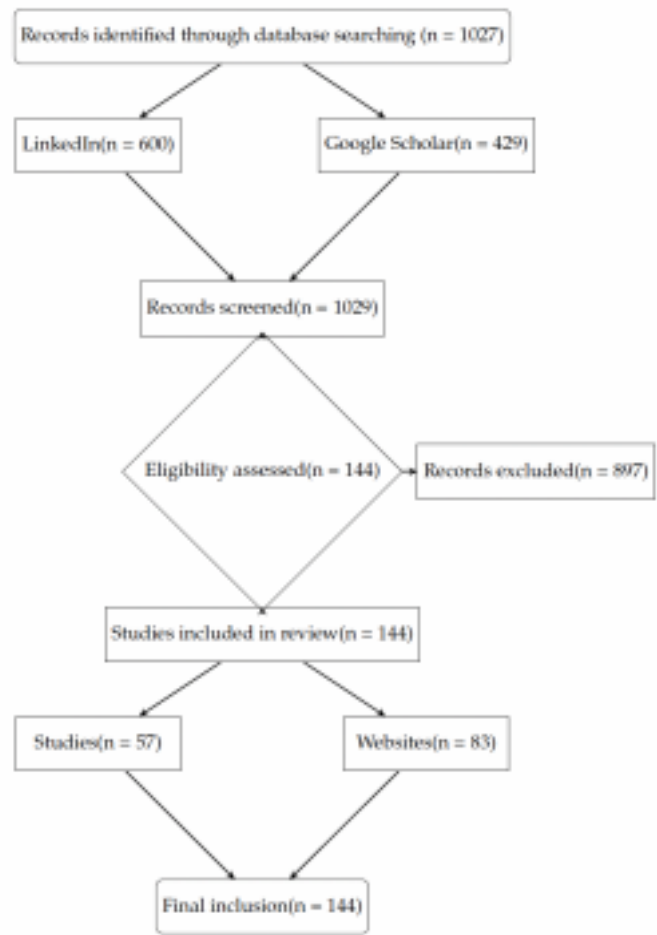

Fig. 1 Flow diagram of the systematic literature review process [31]

# 4 Results

The systematic review included articles from both white and grey literature spanning from 2018 to 2024. We conducted topic modeling on article titles using Latent Dirichlet Allocation (LDA) as described by Blei et al.. The analysis was conducted using Python script which was generated with the help of ChatGPT and presented in Appendix section. The Table 3 provides a holistic and objective view of research data based on LDA topic modelling of titles.

In the following subsections, we present the key themes developed during the anal ysis , detailing the critical factors and dimensions that emerged . We begin with the Metaverse User Experience Maturity Model, followed by the Metaverse Business Readiness Model, and conclude with the Metaverse National Competitiveness Model.

| Topic | Keywords |
|---|---|
| Topic 1 | metaverse, business, engineering, human, security, survey, time, make, acces sible, enters |
| Topic 2 | metaverse, virtual, future, world, new, proceedings, international, physical,computer |
| Topic 3 | metaverse, user, digital, experience, ux, xr, design, fashion, brands, creation |
| Topic 4 | virtual reality, internet, systems, reducing, industrial, things, biggest, acces sibility, motion |
| Topic 5 | metaverse, open, challenges, future, research, practice, state, social, applica tions, vr |

Table 3 LDA Topic Modeling

## 4.1 RQ1:What are the primary elements affecting individual users' adoption and utilization of the metaverse?

The responses to this question have been organized into the main themes outlined in Table 4, while Figure 2 presents the conceptual model that addresses this research question.

Table 4 Thematic Analysis for RQ1: User Experience Factors

| Open Coding | Axial Coding | Selective Coding | Example Articles |
|---|---|---|---|

| Motion sickness, Motorstrain, Hand fatigue, Visual impairments | Barriers to use | Accessibility | [45–48] |
|---|---|---|---|
| Multiple input methods,Customization options, Flexible onboarding, Navigation options | User adaptability | Flexibility | [49–52] |
| Game mechanics,Storytelling, Virtual economies, Interactivity, Presence | Game-like elements | Gamification | [50, 53, 54] |
| Cross-platform artifacts,Data continuity, Open standards | Seamless transitions | Interoperability | [51, 55–57] |
| Self-representation, Non verbal communication, Multi-presence, Multi identity | User interactions | Social | [51, 58, 59] |
| Privacy controls, Security settings, Virtual asset ownership, Content creation, Data trans parency | User control | User Empowerment | [55, 60, 61] |

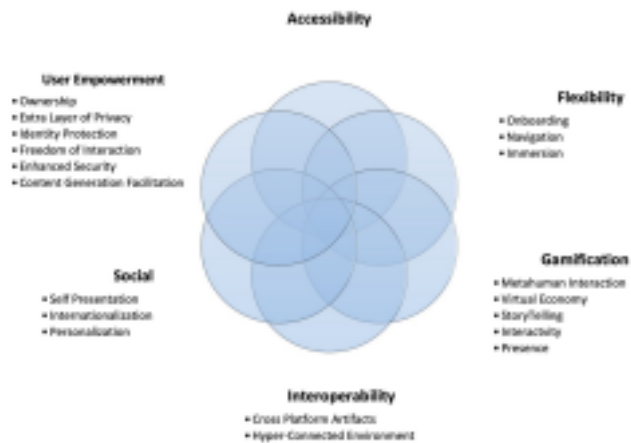

Fig. 2 User Experience Model

### 4.1.1 Accessibility

Accessibility principles for non-immersive environments have been developed over many years; however, the metaverse presents both new opportunities and challenges in this regard. On one hand, the metaverse allows users to participate in events without the limitations of geographic boundaries [62–64]. On the other hand, barriers such as motion sickness from VR headsets, motor strain, and hand fatigue may hinder users from fully engaging with the metaverse [45, 47–50, 65].

One of the major challenges is the lack of agreed-upon criteria for developing third party accessibility tools and guidelines for the metaverse [4, 66]. To address this issue, the World Wide Web Consortium (W3C) has developed XR Accessibility User Requirements, which aims to create a common standard for accessibility in the meta verse [67]. This document considered the topics such as immersive personalization for people with disabilities, and encourage avoiding interactions that trigger motion sickness. We should also note that metaverse projects are not inclusive for people with visual impairments yet [50, 64]. It is encouraging to see projects like Microsoft Canetroller [68], which allows blind people to explore virtual realities. However, more work is needed to ensure that the metaverse is accessible to all users, regardless of their abilities. By developing and adhering to accessibility guidelines, we can create a more inclusive and equitable metaverse that everyone can enjoy, and have a voice. Despite the considerable attention the research community has given to accessibility in the XR field, there is still a lack of sufficient and equitable research on potential immersive threats that may impact neurodiverse users in XR environments [69]. This gap under scores the importance of designing customizable and flexible immersive environments to support individuals with cognitive differences, such as autism, ADHD, and dyslexia. XR technologies have the potential to reduce sensory overload, enhance focus, and provide personalized social interactions, making virtual experiences more inclusive. By prioritizing accessibility and personalization, XR can create spaces where neurodiverse individuals can fully engage, interact, and thrive in immersive environments.

### 4.1.2 Flexibility

User experience studies emphasize the importance of flexibility in allowing users to complete tasks using a variety of methods [70] by delivering multimodal, multi task, and embodied interactions [71]. This principle also necessary for the metaverse ser vices [51, 72], where users should be able to interact with virtual environments in flexible ways, just as they can in the real world. Designing metaverse applications with flexibility in mind can also enhance accessibility. However, there are common barriers to flexibility in the metaverse, including inflexible onboarding, navigation, and immer sion, as well as the need to set up blockchain-based wallets [49–51, 73, 74] . Users have criticized the process of signing up for metaverse apps, and some have experienced fatigue from using XR-based metaverse systems, which rely heavily on reality-based interactions and may require users to stand up [65, 75]. Flexibility in the level of immersion refers to a system's ability to allow users to change their experience along the virtuality continuum [76]. Recent years have seen the introduction of reality-aware VR headsets, which have improved this factor [77].

### 4.1.3 Gamification

Although the applications of the metaverse extend far beyond gaming and entertainment, the principles of game design and elements of gamification play a crucial role in the user experience of metaverse services [50, 51, 53, 54, 78]. The majority of meta verse applications such as Roblox and Horizon World rely on 3D rendering and 3D game engines [50, 79, 80]. Even 2D metaverse applications such as Gathertown [80] are made of gaming elements such as game maps. As a result, it might make more sense to shift from using the term users to using the term players instead, since this new mindset can help us to consider gamification elements in the design of the metaverse services [81].

Game economies are the important game mechanics and they shaped the economy of metaverse. The economy of virtual worlds, and the rising importance of virtual goods, together with proof of ownership with blockchain technology are creating meta verse commerce [53, 55]. Metaverse commerce relies on the convergence of physical and virtual world economies, where the people buy physical goods in virtual immer sive worlds, and buy virtual goods in physical stores [82, 83]. Metaverse experience designers also should expand their storytelling skill by inspiring from narrative and game experience authoring in video games [79, 83]. Storytelling in virtual realities has enabled researchers to design experiences for learners to teach complex topics such as immunology [84]. Metaverse experience designers can move beyond virtual world in storytelling, and they can utilize physical objects and locations for story telling too [85, 86].

Interaction with metaverse services is more than human-human interaction, and the agent personas (meta humans) are essential part of social interactions in meta verse [71]. Interaction with and between metahumans is part of an experience that is not achievable in the real world [87]. Metahuman refers to a virtual image that relies on artificial intelligence technology to imitate humans [88]. In the metaverse, people can create personal digital twins, and use them for interaction with other peo ple, or with the help of personal cognitive digital twins, they can be in more than one place [57, 89]. Also, in the metaverse people may interact with AIs that do not represent a real-world person [72, 90]. Although bringing game-like interactivity to non-game purposes is not a new concept [91–93], the immersive nature of the meta verse increases its importance, and metaverse applications should provide some level of interactive responsiveness [73, 94, 95]. In this context, we use the definition of [96] which defines interactivity as the degree to which users of a medium can influence the form or content of the mediated environment.

Since the interaction with metaverse services mostly happen in virtual or aug mented spaces, presence is another aspect from game studies that shapes the user experience [73, 97]. Presence in virtual reality is mostly about the psychological sense of being in a virtual environment, while in augmented reality is about the level of feel ing that virtual objects belong to the real world [98, 99]. Thus, the metaverse, with its ability to offer heightened realism and a sense of presence, has the potential to become a promising platform that meets the social needs of young people, addressing the gap left by the decline in public events and face-to-face interactions.

Gamification of the metaverse experience was found conceptually interconnected with user empowerment; unlike conventional gamification, which might be control ling the user, gamification in metaverse gives back the freedom and autonomy to the users[100].

### 4.1.4 Interoperability

Interoperability is a widely recognized criterion for the success of the metaverse [50, 51, 55, 57, 101–105]. It allows users to seamlessly transition between different platforms and virtual worlds, without being restricted to a single platform. With inter operability, users can keep access to their virtual belongings across different domains and boundaries of active experiences, making it easier to continue the same experi ence across different virtual worlds [55, 102, 106]. To achieve interoperability, modular and open standards are needed to facilitate the storage and retrieval of cross-platform artifacts and data continuity across different worlds [55, 102, 106]. Cross-platform artifacts enable users to bring entities to multiple virtual worlds. For example, a user can purchase a physical outfit from a physical vendor and the virtual representation of that outfit can be transferred to the user's wallet. This outfit can then be used in a 2D pixelated world like Gathertown, as well as in a high-fidelity environment like Grand Theft Auto Online. To achieve this goal, protocols that facilitate multi-fidelity entities are necessary, and designing multi-fidelity artifacts can become an essential skill for future designers. Also, there should be interfaces to establish hyperconnected environments [107] which connect physical worlds to metaverse environment.

### 4.1.5 Social

Social interactions are a crucial component of the Metaverse [46, 52]. In the future, the social dimension of the metaverse is anticipated to greatly transform how people connect, collaborate, and form relationships within virtual environments. These Inter actions require personalized self-representation to facilitate effective communication and collaboration [51, 58].

The Metaverse must allow users to personalize their metahumans to reflect their social intentions, which could include living with multiple identities simultaneously [57, 59]. Effective social communication in the Metaverse requires non-verbal commu nication such as facial expressions and body movements, which should be incorporated into Metaverse services [108, 109]. Additionally, internationalization is an important aspect of social interactions, and advances in technology such as machine-enabled translation can enable people with different languages to communicate [110]. As communication latency decreases, people from different physical locations can attend live events together and interact with each other in real-time [111, 112]. The capa bility to have multi-presence and multi-identity can have significant impacts on the social user experience and design patterns, which may have unforeseen sociocultural consequences.

### 4.1.6 User Empowerment

The vision of the metaverse is to provide users with more control over their experiences and more leeway to make their own decisions, in contrast to the online services that are currently the status quo, and which bind users to the policies of walled gardens. All of the different aspects that we have discussed up until this point, in addition to the safety and privacy concerns that have been brought up, such as identity protection [60] and self-sovereign identity [55], implicitly place an emphasis on the idea of user empowerment. Interoperability fosters increased user control, which in turn fosters greater autonomy.

Decentralization is a key principle in achieving user empowerment in the meta verse. It aims to reduce the control of service providers over user experiences and data, potentially enhancing security and privacy. However, the implementation of decentral ized systems can sometimes create new challenges for users. For instance, decentralized metaverse applications like Decentraland require users to set up blockchain-based wal lets, which can be a complex and confusing process [49, 113]. This highlights the tension between empowering users with greater control and maintaining ease of use, a balance that metaverse designers must carefully consider.

In order to helps users to have meaningful experiences within the metaverse, the designers need to provide users with the means to determine their own privacy and security preferences, as well as give users control over their data [61, 114].The meta verse presents four overarching categories of security and privacy issues, corresponding to its key characteristics: socialization, immersive interaction, real world-building, and expandability [115]. Socialization-related issues encompass network security vulnera bilities such as injection attacks, man-in-the-middle attacks, and Cross-Site Scripting, as well as privacy leakage from user profiles and data breaches.

Immersive interaction challenges involve risks associated with device communi cation, including insecure deserialization and the potential compromise of biometric data from wearable devices and sensors. Real world-building aspects raise concerns about user profiling and privacy, when user interact with the virtual equivalent of real world systems. Lastly, expandability issues stem from the interconnected nature of the metaverse, creating opportunities for third-party tracking and cross-app tracking, potentially leading to unauthorized access to system components and app information [115].

In addition to this, it encourages the ownership of property rights and virtual goods. The Metaverse paradigm recognizes property rights in virtual worlds, in con trast to the current standard of game economies, in which companies can manipulate the virtual properties of users and they may forbid the trade of virtual goods with real money like in World of Warcraft [51, 55, 61]. In this way, decentralization is a key governing principle [55] and would make the users virtual ownerships persistent [94] and reduce the control of service providers over the the experiences. Decentralization, makes the users shareholder of the service, and rules can be changed only by the decisions of the community [66].

Because the gamified nature of the metaverse encourages freedom of interaction, designers of experiences within the metaverse need to take into consideration a greater variety of interactive variables when developing their concepts. Additionally, they should make use of features that make the process of content generation easier [95]. The facilitation of content generation not only enables users to express their creativity [61], but also enables users to express who they are [116]. AI tools can be implemented by designers to make the process of content generation for their users easier [90].

Emerging technologies like Brain-Computer Interfaces (BCI) promise to take user empowerment to unprecedented levels, potentially allowing users to control their meta verse experiences and interactions directly through their thoughts, further blurring the line between user intent and virtual action [117].

## 4.2 RQ2: What are the main factors driving the adoption and use of the metaverse within businesses?

In the literature there were discussion about how business organization should be prepared to utilize metaverse to achieve their goals and be prepared for transformation in socio-economic dynamics of business environment. Table 5 details the coding process of the study. As illustrated in Figure 3, and discussed in following paragraphs the organizations should prepare for digital twin integration, change in marketing efforts, new service development, and preparing both human and infrastructure resources for transition toward metaverse.

### 4.2.1 Digital Twin Integration

Digital twins serve as a critical bridge between physical and virtual realities, enabling bidirectional information exchange between virtual world and physical world to deliver seamless interactive metaverse environments[128]. Digital twins are digital represen tations of real-world entities or systems [128, 129], and it's crucial to businesses that their processes, assets, and infrastructure are represented and integrated by digital twins. Digital twins facilitate the integration of digital world with the real economy, and can create a seamless user experience and bring efficiency to the business process [119, 120, 128].

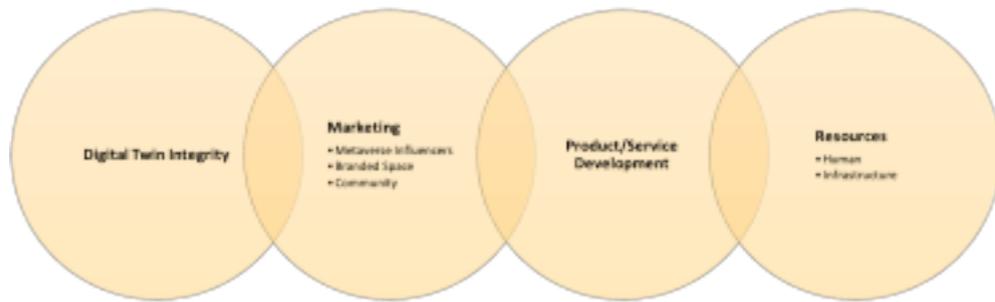

Fig. 3 Business Readiness Model

Table 5 Thematic Analysis for RQ2: Business Readiness Factors

| Open Coding | Axial Coding | Selective Coding | Example Articles |
|---|---|---|---|
| 3D modeling, IoT integration, Virtual represen tations | Real-world integration | Digital Twin Integration | [118–120] |
| Immersive brandexperiences, Virtual influencers, Community building, Virtual retail | New marketing channels | Marketing | [120–123] |
| Virtual product testing,Customer behavior data, Pure virtual ventures, Hybrid businesses | Innovation facilitation | Product/Service Development | [122, 124, 125] |
| VR/AR hardware,Processing systems, Employee training, New job roles | Capability building | Resources | [122, 126, 127] |

For instance, a fashion design company may need to develop a 3D model of their store and staff, or a manufacturer may need to create a digital twin of their factory management system and connect the facilities to them via Internet of Things. The ability of a company to create a digital twin of its business and integrate it with the real world can be an important factor in a company's readiness for the metaverse.

Digital twin, by integrating real-world data such as sensing that, businesses can experiment in the virtual world, and one example is using the virtual world for training the automating training driving algorithms and testing vehicles [130, 131].

## 4.2.2 Marketing

Metaverse provides a new channel for marketing that goes beyond social media, email, web, mobile and search engine presence, and creates a more immersive style of com municating with the brand's audience than ever before [120–122, 126]. People in the Metaverse may not only interact with digital twins and avatars of real people, but may also interact with characters who were created specifically for virtual realities and are representative of real people. As a result, brands aren't just competing for the right to recruit virtual avatars of real celebrities such as Kendal Jenner. Marketers have also become more competitive in recruiting meta influencers that aren't representa tive of actual people anymore, a situation that has changed in the past few years. For instance, Kyra, an Indian meta influencer, has a following of more than 200K from Gen Z on Instagram [132, 133], even though she doesn't represent any actual person. However, KFC and other brands with mascots also have opportunities. By utilizing artificial intelligence, they can use their mascots to interact with more users in a more immersive way [134].

Metaverse is a new social platform for brands [120, 122, 123]. It offers excel lent communication tools and enables brand community members to connect more effectively than on internet blogs and chat rooms. Community growth is a key suc cess criterion for metaverse branding [114, 118] , brands can mint their own tokens, and virtual properties in the metaverse, and community growth is the key factor in generating value for these tokens and brand equity.

The metaverse also creates a new dimension in retail. Previously, companies shopped online, or in brick-and-mortar stores. The metaverse adds a new layer to both of these established paradigms. As a consequence, companies can create new shopping experiences and immersive content for virtual worlds of various themes and graphics [119, 120, 135]. It is possible for customers to visit virtual expos, or to look for merchandise in virtual malls together with other shoppers [136, 137]. For exam ple, Gucci created a Gucci garden on the Roblox platform, and customers could buy digital outfits on this platform [138]. But this new medium is not limited to virtual goods and users can shop for physical goods in the metaverse too. In addition, IoT and Augmented Reality can disrupt the traditional paradigm of sales and customer conversion by integrating physical, branded or non-branded spaces with IoT and Aug mented Reality [107]. Customers can try cosmetic products with augmented reality at stores [139].

Furthermore, virtual worlds such as Pokemon GO merge the physical and vir tual worlds, creating new opportunities for marketing and shopping. It is possible for retailers to attract Gen Z by housing in-game items with their physical stores, and by encouraging them to visit them physically [140].

## 4.2.3 Product/Service Development

The metaverse offers opportunities for creating new types of products and services as well as new ways of product development. A company's ability to facilitate innovation in response to the new realities brought about by the metaverse determines if it is prepared for the economic consequences of the metaverse. The metaverse can help designers conduct design research by allowing them to test unbuilt products in virtual environments [122, 124] and gather data on customers behaviour [141] .Moreover, The metaverse enables the formation of both pure virtual and hybrid business activities [124].

Pure virtual ventures can utilize the metaverse to provide unique products and services that are only available in virtual realities [141]. For example, in Second Life, people can enjoy virtual tourism by visiting worlds that do not exist in the real world and enjoying their destination by meeting imaginary creatures such as vampires [142, 143].

Hybrid businesses are based on the merging of virtual realities with the physical world; Disney is a good example of this disruption, enlivening the creatures of its fiction in its cruise lines. Disney created an AR app that allows cruise passengers to travel with virtual creatures on cruise ships.

The extent to which a company uses metaverse in product development can indicate its readiness for transformation by metaverse. The use of VR in the product development process is a good indicator. Volvo has used the Unity game engine and VR headset to simulate its cars in virtual tracks, reducing physical prototyping and speeding up design iteration [131].

### 4.2.4 Resources

Preparing for the metaverse requires companies to invest in upgrading their technology infrastructure and human resources to cope with an increasingly complex environment [122, 126, 127]. This means not only investing in technical hardware, such as VR/AR head-mounted displays and more powerful processing systems, but also training employees in the skills necessary to apply the metaverse to their everyday workflow.

Immersive design, storytelling, and programming are just a few examples of the skills that will become increasingly important as the metaverse evolves [95]. As part of their infrastructure investments, some companies, such as Accenture, have begun providing their employees with XR head-mounted displays. As a result of this investment, the company can take advantage of the metaverse for corporate training, as well as preparing the mindset of employees for future developments [144]. Indeed, defining metaverse initiatives and creating new roles will also be an important aspect of preparing for the metaverse. As the metaverse evolves, companies will need to have dedicated personnel who are responsible for planning and executing metaverse-related projects. This may include creating new job titles such as Chief Metaverse Officer [145], World Builder, Avatar Clothing Designers, and Metaverse Event Directors.

Metaverse Event Directors will be responsible for planning and executing virtual events within the metaverse. This may include concerts, conferences, gaming events and other types of events that are designed to engage users within the virtual envi ronment. They will work closely with other departments to ensure that all events are aligned with the company's overall goals and objectives [146].

Avatar clothing designers will be responsible for designing and creating virtual clothing for avatars within the metaverse. As such, clothing must be designed with a combination of creativity and technical competence in order to match the fidelity and mechanics of the virtual environment.

Furthermore, as mentioned in the Metaverse User Experience Maturity section, gamification is a core element of metaverse, and regardless of whether a company operates in the entertainment industry, they still need positions like world builders, which were previously found only in video game companies [147]. By creating these new roles and hiring personnel with the necessary skills and expertise, companies can position themselves for success in the metaverse. Companies should also continue to monitor the evolving metaverse and adapt their strategies and initiatives accordingly.

## 4.3 RQ3: What are the significant determinants impacting the adoption and utilization of the metaverse at the national level?

Metaverse poses new competition and opportunities challenges. The researchers utilize the World Economic Forum's terminologies of network readiness and competitiveness to for selective coding of country level factors. This refers to the various drivers that determine a country's level of capability in utilizing the benefits of transitioning to the metaverse [148, 149]. The coding processes is presented in table ??.

Based on these coding, the metaverse national competitiveness model was devel oped which is presented in Figure 4 and its key elements which are the main themes of thematic analysis are presented in below paragraphs:

Table 6 Thematic Analysis for RQ3: National Competitiveness Factors

| Open Coding | Axial Coding | Selective Coding | Example Articles |
|---|---|---|---|
| Infrastructure invest ment, Trade legislation, Financing, Standards, Regulations | Governance framework | Policy | [150–153] |
| Public services in meta verse, Virtual economy size, Sector diversifica tion, Innovation metrics | Economic development | Service Develop ment | [152, 154–156] |

| AR/VR device access,Workforce skills, Educa tion and training | Workforce readiness | Human CapitalReadiness | [157–159] |
| Cultural export, Vir tual citizenship, Virtual nations | Cultural influence | Sociocultural | [152, 160, 161] |

## 4.3.1 Policy

Many governments, including China, the United Arab Emirates, Saudi Arabia, and South Korea, view the metaverse as a critical competition arena and are eager to leverage it to advance their nations in this new era as a new battleground [52, 150–153]. To enhance the competitiveness of their economies in the metaverse, these governments recognize that policies related to infrastructure investment, trade legislation, financing, and standards are essential [162, 163].

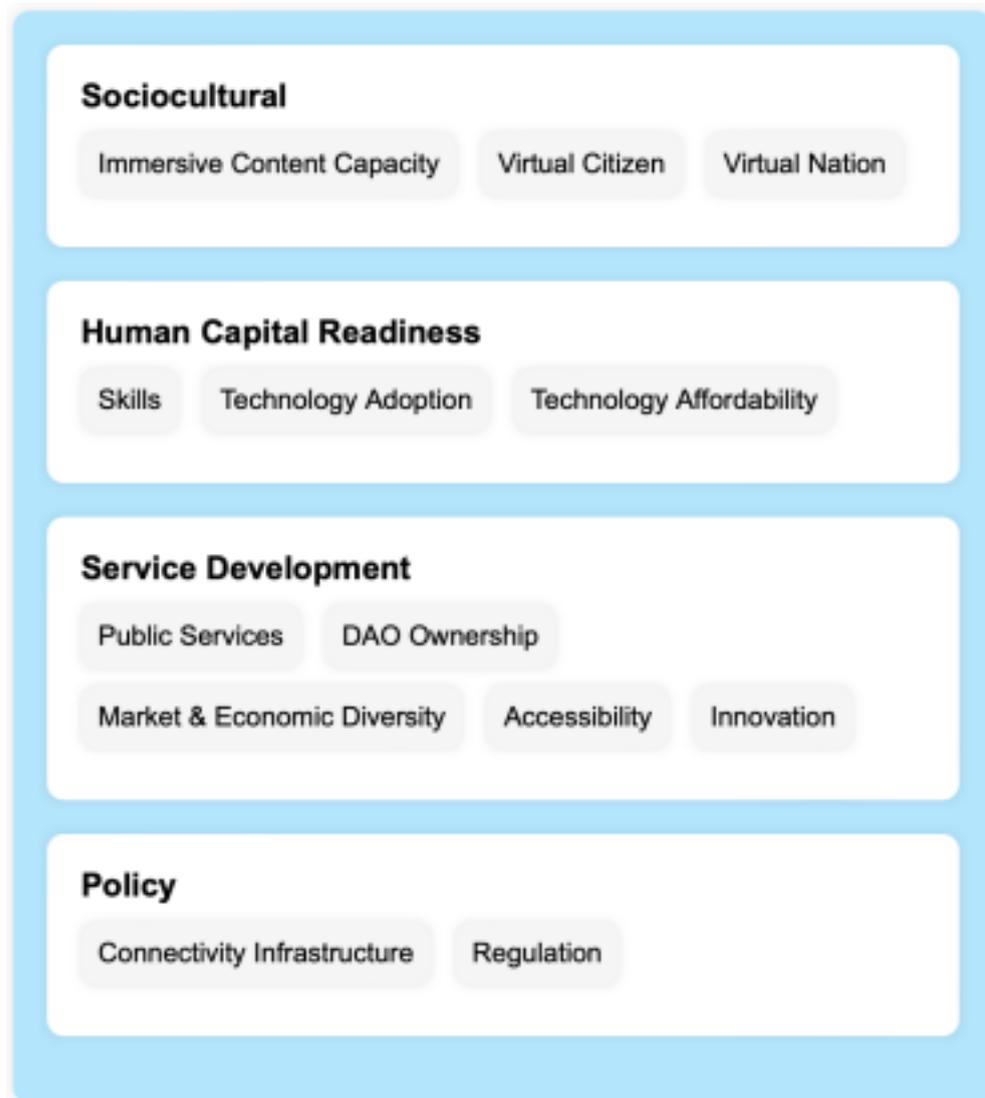

Fig. 4 Metaverse National Competitiveness Model

Building a metaverse requires an infrastructure that enables hyperconnectivity, with 5G and 6G being essential requirements to achieve this goal [14, 128, 157, 164, 165]. While governments aim to attract investment to expand 5G coverage in their countries, they also need to consider physical infrastructure ownership, which can impact the security of their networks [166]. 6G connectivity enables businesses to better more effective digital twin systems [165] which is discussed as an important factors for the readiness of businesses to adapt metaverse. In addition, it is necessary for real time delivery of 3D content, and stability of interactions, and minimizing the latency which is the enemy of metaverse [167].

In addition to this, the new world of the metaverse demands new regulations to safeguard consumers [168] and investors [169]. Innovation requires investment, and investors seek countries with lower investment risks. Metaverse projects and services necessitate regulations that facilitate new business models. However, digital twins and decentralization generate new legal and monetary issues, which place some innovations in the legal gray areas.

This regularity uncertainty creates additional risk for investment in meta verse start-ups. Also, lawmakers needs to consider the emergence of decentralized autonomous organizations. DAOs are analogous to the government approved corporate entities, and the relations between owners are managed by smart contracts. A DAO is community that is cooperatively owned and governed by their members, and man aged by a set of rules enforced on a blockchain by majority consensus [108, 170, 171] , such as proof of stake, and proof of work.

For the regulation of the metaverse it is important to note that, the overreach of government can impact the user empowerment. Aggressive regulation can limit the freedom of speech [172, 173], which may impact the metaverse service providers [174].

### 4.3.2 Service Development

Metaverse will transforms the interaction with services, and create opportunities for creation of completely new services. Government can utilize metaverse to create to public services to citizens, and improve their productivity [152, 154] . For instance, Barbados has opened a diplomatic embassy in Decentraland , a premier Metaverse platform [155]. South Korea also started a five-year plan to build the digital twin of its capital city and provide city services in metaverse [156].

The investment in metaverse and the size of virtual economy is another indicator for the status of service development in countries. Countries are investing billions of dollars to create business environment for creation of metaverse companies, expand their metaverse market size, and gain economic hegemony in coming years [152, 175]. Also, China hopes a public-private partnership bring the authoritarian values of CCP to the metaverse, and strictly regulate the metaverse to control the public opinions [176].

Also, it is important to note that metaverse creates new opportunities in various sectors, and countries should diversify the investment in metaverse market since it can disrupt various sectors namely as tourism, financial services, entertainment, education, shopping, and even healthcare [177–181].

The current large online platforms are controlled by specific private companies, and regulatory agencies can control this firms, and enforce them by legal orders. European Union use the terminology of gatekeeper to describe "digital platforms that provide an important gateway which grant them the power to act as a private rule maker"[

European Commission], but the common perception is that metaverse relies on decen tralization. In a decentralized metaverse, platforms will not be owned and managed by a central gatekeeper, but they managed by DAOs [182].

In other word, the gatekeepers of the decentralized metaverse platforms would a large community of people, and individuals across the worlds can make rules for platforms and ecosystems by their share of processing infrastructure and tokens, and it imply that countries with who owns more blockchain nodes have more voting power [170].

The extent of innovation can define the status of competitiveness and governments should set right conditions to incentivize innovation [178]. Patents are one of common indicators for measuring innovation [183], and currently, United States, China and South Korea are leading countries by patents for metaverse technologies [184].

Ensuring accessibility and inclusion is a critical factor for the success and widespread adoption of metaverse technologies. As the metaverse becomes more inte grated into our daily lives, it is important that it is accessible to people of all abilities, including those with disabilities. Accessibility and inclusion should be at the forefront of policymaking efforts to make metaverse accessible to everyone [157, 158, 178], and they can also drive innovation, so policy makers should plan accordingly to make metaverse accessible in their countries [181].

### 4.3.3 Human Capital Readiness

Populations with greater access to metaverse technologies, particularly AR/VR devices, will likely have a competitive advantage in the emerging digital landscape. However, the high cost of this equipment can be a barrier to widespread adoption. To address this, institutions and companies may need to explore alternative ways of making these technologies more accessible and affordable.

Additionally, it is important for institutions to invest in building a workforce that is equipped with the skills and knowledge necessary to effectively utilize AR/VR solutions within the context of the metaverse. The essential skills for metaverse are fast-evolving which is make this activity more challenging [181]. This requires a proactive approach to training and development, as well as a commitment to ongo ing education and upskilling. Ultimately, the extent to which a nation can effectively leverage the potential of the metaverse will depend on its ability to develop a human capital base that is ready and able to navigate this new digital landscape [157, 158].
.

### 4.3.4 Sociocultural

The metaverse is expected to have a significant social impact, leading to major cultural and social changes [152]. It will allow people to live in the digital twins of distant cities with different cultures and social values, creating virtual territories for those with shared values. These virtual nations could have profound implications for the geo-political landscape and the alliances that currently hold our nations together [160].

Estonia's e-residency program, which allows individuals outside of Estonia to apply for residency and enjoy the benefits of Estonian citizenship, demonstrates the poten tial for virtual citizenship. With the metaverse, individuals could live in one country while spending time in the digital twin of another. This has the potential to disrupt current perceptions of populations, with countries vying to attract virtual citizens and creating economic zones for metaverse workers [161] which relies on post nationalism interactions [71].

Virtual residency may extend beyond living in the digital twins of other countries. Video game worlds, particularly massive multiplayer online role-playing games like World of Warcraft, have shown the plausibility of living in fictional virtual worlds [185]. Combining this with DAOs can create new political realities [186]. In the game Star Atlas, for example, players use game currency to determine governance from local to top level, similar to the structure of local and federal governments. Countries with greater influence in game design and popular culture may be more successful in creating virtual nations since they have better capacity for immersive content creation.

Using the metaverse to export culture can be a national metaverse strategy [152]. Countries with distinct histories and cultural heritage can motivate citizens of other nations to visit the digital twins of their countries.

The other important factor is the potential impact of metaverse on education sys tem, and education institutions. Metaverse can make education system more available [71, 159]. Metaverse relies on capturing data and creating simulated environments, and it create new usecases for simulated based learning.

There are ethical concerns about that how these virtual layers that need be consid ered, specially the risk associated with replacing the perception of the actual reality with incorrect information.

# 5 Discussion

Previous studies highlighted the role of interoperability [7], open systems, and blockchain in metaverse development. This study found multiple factors answering the research questions on metaverse adoption.

At the individual level, key factors are accessibility, flexibility, gamification, interoperability, social aspects, and user empowerment. For businesses, digital twin integration, marketing capabilities, product/service development, and resource readi ness emerged as crucial. At the national level, policy frameworks, service development, human capital readiness, and sociocultural factors were identified as essential for suty ding the competitiveness. These findings expand on existing literature by providing comprehensive and multidicplinary lens for understanding metaverse adoption and readiness.

User experience in the metaverse acts as a catalyst for business innovation and growth by driving adoption, and engagement. As users spend more time in immersive and interactive experiences, businesses can develop novel products, and services tai lored to this new hybrid reality. These innovations, in turn, contribute to economic growth by creating new markets, job opportunities, and revenue streams within the metaverse ecosystem.

The resulting economic activity boosts national competitiveness by increasing GDP, attracting foreign investment of virtual citizens, and positioning countries . Governments play a crucial role in shaping this ecosystem by implementing policies that foster innovation, protect user rights, and promote digital literacy. By invest ing in infrastructure, setting regulatory frameworks, and incentivizing research and development, governments can create an environment where accessible and inclu sive user experiences boosts economic interactions increase the the national economy benefits from the transformative potential of the metaverse. This synergistic relation ship between users, businesses, and government requires investment in education and private-public partnership. In following paragraph, we discuss each aspect at details:

## 5.1 Metaverse User Experience Design

Designers of metaverse experiences face the challenge of creating accessible, flexible, and engaging virtual environments that effectively connect people with information and each other using cutting-edge technologies. These factors align with established principles of user-centered design. As the metaverse evolves, designers will need to consider these factors to create engaging and meaningful experiences that appeal to diverse user populations. The model provides a valuable tool for designers to assess the maturity of their metaverse applications and identify areas for improvement.

The user experience in the metaverse must address the fundamental need to con nect people with information and each other in novel ways. Simultaneously, designers must consider how their experiences facilitate the convergence of community with technology, creating digital spaces where social norms and institutional processes can be translated and potentially transformed.

A critical aspect of user experience in the metaverse is ensuring robust secu rity and privacy protections. As users engage more deeply in virtual environments, the amount and sensitivity of data generated will increase dramatically, necessitating strong privacy safeguards and transparent data management practices. Designers and HCI researchers must integrate these considerations into their user-centered approach to create truly empowering and trustworthy metaverse experiences.

In order to gain public trust and encourage participation in metaverse systems that continuously gather data from users' every interaction - including eye movements, body gestures, and even brain activity - it is essential to design effective strategies to enhance the protection of users' privacy and confidentiality of their personal data, and decentralization and blockchain technology can be recognized as a critical principle in addressing these concerns. By reducing the control of service providers over user experiences and data, decentralized systems can enhance security and privacy.

Based on the recognized factors, the authors developed a heuristic evaluation rubric that still needs to be evaluated. This rburic is presented in Appendix A, and it can guide user experience researchers and interaction designers study their metaverse applications and plan for the future lifecycle of the products.

## 5.2 Metaverse Business Planning

The Metaverse Business Readiness Model identifies key areas where businesses need to focus their efforts to create innovative solutions for connecting customers, products/services, and information using emerging metaverse technologies.

Moreover, businesses must prepare for new ways of engaging with communities and interfacing with social units in the metaverse, as the convergence of community and technology reshapes traditional business-customer relationship

The model emphasizes the importance of digital twin integrity, marketing, pro duct/service development, and resource allocation in preparing for the metaverse economy. This model can guide corporations in the planning for the disruptive effects of metaverse on their organizations.

Based on these factors we propose a set of questions as a metaverse planning self-assesment checklist to guide business managers in their strategy for metaverse transformation which is presented in Table A3. While this tool is not validated by experts, it summarize the findings of the study as the actionable questions to guide future research.

## 5.3 Metaverse Policy Making

The Metaverse National Competitiveness Model provides a framework for assessing a country's readiness to leverage the metaverse for economic and social development. Policymakers must prioritize digital infrastructure investments, particularly in 5G and 6G networks, to enable the hyper-connectivity required for the metaverse. Simul taneously, they should develop adaptive regulatory frameworks that address digital property rights, virtual currencies, and cross-border transactions.

Education and workforce development programs focused on metaverse-related skills are essential to maintain national competitiveness. Policymakers should also consider how to leverage cultural assets in the virtual space and prepare for evolving concepts of citizenship and governance in the metaverse era.

Metaverse requires a multifaceted strategy that foresees the digital twin strategy, ethical factors regulations that protect individual liberties, infrastructure renovation investment, and maintaining innovation opportunities in public service delivery such as novel education systems.

Based on the discussed factors, the researchers come up with several research ques tions which are presented in Appendix C. These questions are not areas guide future research but they can guide policy makers. Table A4 presents a series of strategic questions designed to assist policymakers, lawmakers, and other stakeholders in assess ing readiness. It can guide the development of comprehensive national technology strategies for future digital realities.

## 5.4 Limitations

However, it is important to acknowledge the limitations of this study. As the meta verse is still an emerging phenomenon, the factors identified in the models may evolve over time as new technologies and use cases emerge. In addition, while the glasserian grounded theory [39] helped the researchers in exploring an emerging phenomena, this methodology is inherently interpretive and subject to researcher bias. Although the authors followed established practices for coding and analysis, their own backgrounds, experiences, and assumptions may have influenced the selection and interpretation of the data. Future research could employ multiple coders and assess inter-rater reliability to enhance the trustworthiness of the findings.

# 6 Conclusion & Future Works

We have reviewed different perspectives on the essential characteristics for building the metaverse based on the works of both industry and academia to develop conceptual models that describe factors that contribute to adaption of metaverse at individual, business, and national level. The conceptual models can be used to study the metaverse state and assist designers, business strategists, and policymakers in exploring and building the metaverse bricks. It also can help researchers to connect the variables of these models together and define new domain space of studies.

To further our research, we plan to validate the models by interviewing user experi ence researchers, business consultants, technology policy researchers, and futurists. We hope that our work will serve as a base for future studies in user experience design of the new generation of interactive systems. Extending and validating heuristic rubrics could be an area of research. Researchers can use it to systematically study the matu rity of metaverse solutions and find areas of improvement. The business checklist can be used by researchers to study the readiness of businesses in various sectors. We are particularly interested in using it in expert interviews with business managers in the retail sector. Last but not least, we suggest policy researchers utilize the national metaverse readiness checklist as a starting point for studying the opportunities and challenges that digital realities pose for the economy, culture, and society.

## Abbreviations

Augmented Reality: AR
Extended Reality: XR

Virtual Reality VR
Artificial Intelligence: AI

# Declarations

## Funding

No funding was received for conducting this study.

## Competing interest

The authors have no relevant financial or non-financial interests to disclose.

## Code availability

The code that creates topic modelling is presented in appendix.

## Authors' contributions

All authors contributed to the study conception and design. Material preparation, data collection and analysis were performed by All authors. The first draft of the manuscript was written by Amir Reza Asadi and all authors commented on previous versions of the manuscript. All authors read and approved the final manuscript.

## Data availability

Not Applicable

## Ethics

Not Applicable: This article does not involve any studies conducted with human participants or animals by the authors.

## Acknowledgements

Not Applicable
If any of the sections are not relevant to your manuscript, please include the heading and write 'Not applicable' for that section.

section[Appendix ]"Script"

```
1 import pandas as pd
2 from sklearn . feature_extraction . text import CountVectorizer
3 from sklearn . decomposition import LatentDirichletAllocation
4 from gensim import corpora
5 from gensim . models . ldamodel import LdaModel
6 from nltk . corpus import stopwords
7 from nltk . tokenize import word_tokenize
8 import nltk
9
```

```python
nltk . download ('punkt ')
nltk . download ('stopwords ')

# Replace 'your_file . csv ' with your actual file path
df = pd . read_csv ('your_file . csv ')

# Assuming the article titles are in a column named 'title '
titles = df ['title ']. values

# Function to preprocess text
def preprocess ( text ):
    stop_words = set ( stopwords . words ('english ') )
    tokens = word_tokenize ( text . lower () )
    filtered_tokens = [ word for word in tokens if word . isalpha () and word not in stop_words ]
    return ' '. join ( filtered_tokens )

# Preprocess all titles
preprocessed_titles = [ preprocess ( title ) for title in titles ]

# Create a document - term matrix using CountVectorizer
vectorizer = CountVectorizer ( max_df =0.95 , min_df =2 , stop_words ='english ') dtm = vectorizer .
fit_transform ( preprocessed_titles )

# Set the number of topics
n_topics = 5

# Fit the LDA model
```

```python
lda_model = LatentDirichletAllocation ( n_components = n_topics , random_state =42) lda_model . fit ( dtm )

def display_topics ( model , feature_names , no_top_words ):
    for topic_idx , topic in enumerate ( model . components_ ):
        print (f" Topic { topic_idx + 1}: ")
        print (" ". join ([ feature_names [ i] for i in topic . argsort () [: - no_top_words -
                1: -1]]) )

no_top_words = 10
feature_names = vectorizer . get_feature_names_out ()
display_topics ( lda_model , feature_names , no_top_words )
```

# Appendix A Metaverse Service Heuristic Evaluation

Table A1: Metaverse UX Maturity Heuristic Evaluation Rubric

Heuristic Rating (1- 5)

1. Accessibility
1.1 Supports multiple input methods
1.2 Provides interaction options for users
with visual, auditory, or motor impairments
1.3 Allows customization of sensory output
1.4 Employed W3C XR Accessibility User Requirements

2. Flexibility

2.1 Offers multiple ways to complete tasks
2.2 Provides flexible onboarding process
2.3 Allows users to change level of immersion
2.4 Supports expected navigation methods

3. Gamification
3.1 Incorporates engaging game mechanics
3.2 Utilizes storytelling elements

effectively
3.3 Provides clear goals and progression
3.4 Balances challenge and skill level
3.5 Integrates virtual economy elements

4. Interoperability
4.1 Deliver cross-platform service
4.2 Allows transition between different vir
tual environments

26
Notes

Table A1 continued from previous page

4.3 Enables persistence of user-owned vir
tual items across platforms
4.4 Utilizes open standards for

Heuristic Rating (1- 5)

data and
asset portability
4.5 Supports multi-fidelity entities

5. Social
5.1 Provides self presentation customiza
tion options
5.2 Supports non-verbal communication
such as facial expressions
5.3 Enables multi-presence and multi
identity capabilities
5.4 Provide options for cross-language com
munication

6. User Empowerment
6.1 Provides privacy and security control
settings

6.2 Allows users to own and manage their
virtual assets
6.3 Supports user-generated content cre
ation
6.4 Enables community-driven governance
(if applicable)
6.5 Offers transparency in data collection
6.6 System addresses the privacy concerns
of bystanders

Table A2 Overall Evaluation

Notes

| Category | Average Rating | Strengths | Areas for Improvement |
|---|---|---|---|
| Accessibility | | | |
| Flexibility | | | |
| Gamification | | | |
| Interoperability | | | |
| Social | | | |
| User Empowerment | | | |

Overall Metaverse UX Maturity Score: / 5
Additional Comments:

Instructions: Rate each heuristic on a scale of 1-5, where:

- 1 = Poor implementation
- 2 = Basic implementation
- 3 = Adequate implementation
- 4 = Good implementation

• 5 = Excellent implementation

# Appendix B Metaverse Business Readiness Self Assesment Checklist

Table B3: Metaverse Business Readiness Self Assesment Checklist

Assessment Item Status Digital Twin Integration

☐

Have we identified key business processes that could benefit from digital twin integration?

☐

Do we have the technical capabilities to create the digital twins of our business?

Marketing

Have we explored metaverse marketing channels ? ☐ Are we prepared to create and manage virtual brand experiences ? ☐ Do we considered an strategy for virtual brand ambassadors? ☐

☐

Do we have a strategy for community building in virtual enviornments?

Product/Service Development

☐

Have we identified opportunities to create new virtual products or services?

☐

Are we using virtual environments for product testing or customer feedback?

☐

Have we considered how our existing products or services could be adapted for the metaverse?

☐

Do we have processes in place to facilitate innovation in virtual environments?

☐

Have we studied the potential threats & of metaverse for our services and products?

Resources - Human Resources

☐

Have we identified the skills needed for metaverse-related roles such as 3D designers, virtual event planners?

☐

Do we have a plan for training existing staff on metaverse technologies and concepts?

Table B3: Metaverse Business Readiness Self Assesment Checklist

Assessment Item Status

☐

Are we prepared to create new roles specific to metaverse initiatives (e.g., Chief Metaverse Officer)?

Resources - Infrastructure

☐

Have we assessed our current technological infrastructure's readiness for metaverse integration?

☐

Do we have a plan for investing in essential hardware ( VR/AR devices) and software?

☐

Have we considered the data management and security implications of operating in the metaverse?

General Strategy

☐

Do we have a roadmap for gradually implementing metaverse technologies

# Appendix C National Metaverse Readines Table

C4: National Metaverse Readiness Checklist

| Category | Checklist Questions |
|---|---|
| Infrastructure | • Are investments being made in advanced commu nication technologies?<br>• Are there measures to ensure digital infrastruc ture security and resilience?<br>• What is the state of access to extended reality technology?<br>• What is the state of deployment of digital twin within the infrastructure? |

Table C4 continued from previous page

| Category | Checklist Questions |
|---|---|
| Regulatory and Gover nance Framework | • Do existing policies facilitate research & develop ment of emerging technologies?<br>• Do existing policies accommodate new technology-enabled business models?<br>• Is there a framework for novel forms of digital organization and governance?<br>• Are there initiatives encouraging technological innovations for extended realities?<br>• Is there a framework for intellectual property protection in digital environments? |
| Economic Development | • Are there incentives for innovation in new digital realms?<br>• Is there a strategy to expand presence in emerging digital markets? |
| Accessibility and Inclu sion | • Are there accessibility policies that apply to extended realities?<br>• Are there initiatives to bridge potential digital divides? |

| Skills Development | • Are future-focused digital skills being taught in education/training programs?<br>• Are there programs for continuous workforce reskilling for digital realities?<br>• To what degree does the education system utilize extended reality to advance educational goals? |
|---|---|
| Cultural and Social Fac tors | • Are there initiatives for cultural heritage repre sentation in extended reality?<br>• Are there guidelines for ethical behavior in digital spaces? |

Table C4 continued from previous page

| Category | Checklist Questions |
|---|---|
| Content Creation andCreative Industries | • Are there support mechanisms for digital content creators and industries?<br>• Are there initiatives promoting creative output on emerging platforms? |
| Social Impact | • Have the social impacts of new digital technolo gies been considered? |
| Public Service | • To what degree is the public service sector ready to be delivered via Metaverse?<br>• To what degree is the digital twins are integrated into public service delivery? |

| Global Engagement | • Are attractive digital environments being created for international participation?<br>• What is the proportion of foreign users partici pating in metaverse platforms of a country? • To what extent are we positioned to influence the development and adoption of decentralized standards and Decentralized Autonomous Orga nizations (DAOs)? For example, what percentage of global hash power in major blockchain net works is contributed by our country's mining operations?<br>• Are there strategies for participating in interna tional digital standard-setting?<br>• Are there initiatives promoting cross-border dig ital collaboration?<br>• What is the level of our involvement in developing and proposing decentralized standards? |
|---|---|

# References

[1] Stephenson, N.: Snow Crash. Penguin UK, ??? (1994)

[2] Zhang, G., Cao, J., Liu, D., Qi, J.: Popularity of the metaverse: Embodied social presence theory perspective. Frontiers in psychology 13, 997751 (2022)

[3] Riccitiello, J.: Getting Beyond the Fiction and Seeing the Reality in the Meta verse (2022). https://www.youtube.com/watch?v=hnWzVN42uP0 Accessed 2022-09-12

[4] EI, C.H., TAN, D.: Refreshing Singapore's Social Compact Through Citizen Engagement (2022)

[5] Chen, B.-J., Yang, D.-N.: User Recommendation in Social Metaverse with VR. In: Proceedings of the 31st ACM International Conference on Information & Knowledge Management, pp. 148–158 (2022)

[6] Gallon, R., Lorenzo, N.: From Metastudies to Metaverse: Disrupting the University. International Journal of E-Learning & Distance Education (2023)

[7] Radoff, J.: The Metaverse Value-Chain (2021). https://medium.com/ building-the-metaverse/the-metaverse-value-chain-afcf09e3a7

[8] Digitopia: We are Digitopia! We are all about Business Impact! (2022). https://digitopia.co/uploads/202207 digitopia mri v33.pdf

[9] Ganj, A., Zhao, Y., Su, H., Guo, T.: Mobile ar depth estimation: Challenges & prospects. In: Proceedings of the 25th International Workshop on Mobile Computing Systems and Applications, pp. 21–26 (2024)

[10] Lim, C.H., Lee, S.C.: The effects of degrees of freedom and field of view on motion sickness in a virtual reality context. International Journal of Human–Computer Interaction, 1–13 (2023)

[11] Vidak, A., Sapi´c, I.M., Zahtila, K.: A user experience survey of an augmented ˘ reality android application for learning coulomb's law. Physics Education 59(4), 045033 (2024)

[12] Asadi, A.R., Hemadi, R.: Towards mixed reality as the everyday com puting paradigm: Challenges & design recommendations. arXiv preprint arXiv:2402.15974 (2024)

[13] Bibri, S.E.: The social shaping of the metaverse as an alternative to the imag inaries of data-driven smart cities: A study in science, technology, and society. Smart Cities 5(3), 832–874 (2022)

[14] Bibri, S.E.: The social shaping of the metaverse as an alternative to the imagi naries of data-driven smart Cities: A study in science, technology, and society. Smart Cities 5(3), 832–874 (2022). Publisher: MDPI

[15] Jasanoff, S., Kim, S.-H.: Containing the atom: Sociotechnical imaginaries and nuclear power in the United States and South Korea. Minerva 47, 119–146 (2009). Publisher: Springer

[16] Said, H., Zidar, M., Varlioglu, S., Itodo, C.: A framework for the discipline of information technology. In: Proceedings of the 22nd Annual Conference on Information Technology Education, pp. 53–54 (2021)

[17] Dionisio, J.D.N., III, W.G.B., Gilbert, R.: 3D virtual worlds and the metaverse: Current status and future possibilities. ACM Computing Surveys (CSUR) 45(3), 1–38 (2013). Publisher: ACM New York, NY, USA

[18] Montes, G.A.: What is Spatial Computing and How is it Rev olutionizing Our World? (2022). https://www.verses.ai/blogs/what-is-spatial-computing-and-how-is-it-revolutionizing-our-world Accessed 2023-02-15

[19] Saracco, R.: Digital twins: evolution in manufacturing. IEEE Digital Reality (2022)

[20] Chen, Y., Huang, W., Jiang, X., Zhang, T., Wang, Y., Yan, B., Wang, Z., Chen, Q., Xing, Y., Li, D., *et al.*: Ubimeta: A ubiquitous operating system model for metaverse. International Journal of Crowd Science 7(4), 180–189 (2023)

[21] Carroll, J.M., Shih, P.C., Kropczynski, J.: Community informatics as innovation in sociotechnical infrastructures. The Journal of Community Informatics 11(2), 719–733 (2015)

[22] Bigonah, M., Jamshidi, F., Marghitu, D.: Immersive agricultural education: Gamifying learning with augmented reality and virtual reality. In: Cases on Col laborative Experiential Ecological Literacy for Education, pp. 26–76. IGI Global, ??? (2024)

[23] Jamshidi, F., Bigonah, M., Marghitu, D.: Striking a chord through a mixed methods study of music-based learning to leverage music and creativity to bridge the gender gap in computer science. In: Proceedings of the 55th ACM Technical Symposium on Computer Science Education V. 2, pp. 1694–1695 (2024)

[24] Bigonah, M., Jamshidi, F., Pant, A., Marghitu, D.: Work in progress: Grace platform: Enhancing pedagogy with gamified ar and vr in agriculture education. In: 2024 ASEE Annual Conference & Exposition (2024)

[25] Cai, S., Jiao, X., Song, B.: Open another door to education—applications, challenges and perspectives of the educational metaverse. Metaverse 3(1), 12 (2022)

[26] Meccawy, M.: Creating an immersive xr learning experience: A roadmap for educators. Electronics 11(21), 3547 (2022)

[27] Bui, T.X., Sankaran, S., Sebastian, I.M.: A framework for measuring national e-readiness. International Journal of Electronic Business 1(1), 3–22 (2003).

[28] Alfarraj, O., Alkhalaf, S., Nielsen, S., Alghamdi, R.: The Use of Grounded The ory Techniques in IS research. International Journal of Computers 9, 115–124 (2015)

[29] Chapman, L., Plewes, S.: A UX maturity model: effective introduction of UX into organizations. In: Design, User Experience, and Usability. User Experience Design Practice: Third International Conference, DUXU 2014, Held as Part of HCI International 2014, Heraklion, Crete, Greece, June 22-27, 2014, Proceedings, Part IV 3, pp. 12–22. Springer, ??? (2014)

[30] Elaine, E., Simone, S., Kristel, K.: Increasing the UX maturity level of clients:

A study of best practices in an agile environment. Information and Software Technology 154, 107086 (2023). Publisher: Elsevier

[31] Anwar, R.M.I., Azhar, S.: Mixed reality in building construction inspection and monitoring: A systematic review. In: Virtual Worlds, vol. 3, pp. 319–332 (2024). MDPI

[32] Bhanu, A., Sharma, H., Piratla, K., Chalil Madathil, K.: Application of aug mented reality for remote collaborative work in architecture, engineering, and construction–a systematic review. In: Proceedings of the Human Factors and Ergonomics Society Annual Meeting, vol. 66, pp. 1829–1833 (2022). SAGE Publications Sage CA: Los Angeles, CA

[33] Wolfswinkel, J.F., Furtmueller, E., Wilderom, C.P.: Using grounded theory as a method for rigorously reviewing literature. European journal of information systems 22(1), 45–55 (2013). Publisher: Taylor & Francis

[34] Page, M.J., McKenzie, J.E., Bossuyt, P.M., Boutron, I., Hoffmann, T.C., Mul row, C.D., Shamseer, L., Tetzlaff, J.M., Akl, E.A., Brennan, S.E., et al.: The prisma 2020 statement: an updated guideline for reporting systematic reviews. bmj 372 (2021)

[35] Dopfer, K., Foster, J., Potts, J.: Micro-meso-macro. Journal of evolutionary economics 14, 263–279 (2004). Publisher: Springer

[36] Singjai, A., Simhandl, G., Zdun, U.: On the practitioners' understanding of coupling smells—a grey literature based grounded-theory study. Information and Software Technology 134, 106539 (2021)

[37] Garousi, V., Felderer, M., M̈antyl̈a, M.V.: Guidelines for including grey lit erature and conducting multivocal literature reviews in software engineering. Information and software technology 106, 101–121 (2019)

A study of best practices in an agile environment. Information and Software Technology 154, 107086 (2023). Publisher: Elsevier

[31] Anwar, R.M.I., Azhar, S.: Mixed reality in building construction inspection and monitoring: A systematic review. In: Virtual Worlds, vol. 3, pp. 319–332 (2024). MDPI

[32] Bhanu, A., Sharma, H., Piratla, K., Chalil Madathil, K.: Application of aug mented reality for remote collaborative work in architecture, engineering, and construction–a systematic review. In: Proceedings of the Human Factors and Ergonomics Society Annual Meeting, vol. 66, pp. 1829–1833 (2022). SAGE Publications Sage CA: Los Angeles, CA

[33] Wolfswinkel, J.F., Furtmueller, E., Wilderom, C.P.: Using grounded theory as a method for rigorously reviewing literature. European journal of information systems 22(1), 45–55 (2013). Publisher: Taylor & Francis

[34] Page, M.J., McKenzie, J.E., Bossuyt, P.M., Boutron, I., Hoffmann, T.C., Mul row, C.D., Shamseer, L., Tetzlaff, J.M., Akl, E.A., Brennan, S.E., et al.: The prisma 2020 statement: an updated guideline for reporting systematic reviews. bmj 372 (2021)

[35] Dopfer, K., Foster, J., Potts, J.: Micro-meso-macro. Journal of evolutionary economics 14, 263–279 (2004). Publisher: Springer

[36] Singjai, A., Simhandl, G., Zdun, U.: On the practitioners' understanding of coupling smells—a grey literature based grounded-theory study. Information and Software Technology 134, 106539 (2021)

[37] Garousi, V., Felderer, M., M̈antyl̈a, M.V.: Guidelines for including grey lit erature and conducting multivocal literature reviews in software engineering. Information and software technology 106, 101–121 (2019)

[38] Loganathan, J., Ghai, V., Ilaalagan, R., Doumouchtsis, S.K., CHORUS: Vul vodynia: What is available online? a systematic review of information on the internet. Journal of Obstetrics and Gynaecology Research 48(8), 2112–2121 (2022)

[39] Glaser, B.G., Strauss, A.L., Strutzel, E.: The discovery of grounded theory; strategies for qualitative research. Nursing research 17(4), 364 (1968). Publisher: LWW

[40] Remenyi, D.: Grounded Theory: A Reader for Researchers, Students, Faculty and Others. ACPI Publishing, ??? (2013)

[41] Johnson, B., Holness, K., Porter, W., Hernandez, A.: Complex Adaptive Systems of Systems: A Grounded Theory Approach. Grounded Theory Review 17(1) (2018)

[42] White, P., Devitt, F.: Creating Personas from Design Ethnography and Grounded Theory. Journal of Usability Studies 16(3) (2021)

[43] Doyle, D.T., Brubaker, J.R.: Digital legacy: a systematic literature review. Proceedings of the ACM on Human-Computer Interaction 7(CSCW2), 1–26 (2023)

[44] Blei, D.M., Ng, A.Y., Jordan, M.I.: Latent dirichlet allocation. Journal of machine Learning research 3(Jan), 993–1022 (2003)

[45] Eid, N.: Making the Metaverse Accessible to Diversity, Equity, and Inclusion (2022). https://www.ruhglobal.com/making-the-metaverse-accessible-to-diversity-equity-and-inclusion/

[46] SeongJeong Yoon, S J Choi: A Study on the Future Prospect of the Metaverse through Human Cognition 21(7), 149–168 (2023) https://doi.org/10.14801/jkiit.2023.21.7.149 . Chap. 0. MAG ID: 4385689220 — RAYYAN-INCLUSION: "Amir"=¿"Included"

[47] Stoner, G.: VR Is Here to Stay. It's Time to Make It Accessi ble (2022). https://www-wired-com.cdn.ampproject.org/c/s/www.wired.com/story/virtual-reality-accessibility-disabilities/amp Accessed 2023-02-01

[48] Evans, M.: Virtual Reality Accessibility: The Importance of Com fort Ratings and Reducing Motion (2022). https://equalentry.com/virtual-reality-accessibility-comfort-ratings-and-reduced-motion/ Accessed 2023-02-01

[49] Lu, J., Chang, A.: Making the metaverse mainstream is about the user experience. Here's why (2022). https://www.weforum.org/agenda/2022/04/making-metaverse-mainstream-user-experience/ Accessed 2023-02-01

[50] Nicholls, P.: Can Designers Make the Metaverse Less Awkward? (2022). https:// www.toptal.com/designers/ux/designing-for-the-metaverse Accessed 2022-02- 01

[51] Kopinsky, R.: Designing For The Metaverse | 5 UX/UI Considerations (2023). https://www.youtube.com/watch?v=HMFTUQCN8xY

[52] Huansheng Ning, Hang Wang, Yujia Lin, Wenxi Wang, Sahraoui Dhe lim, Fadi Farha, Jianguo Ding, Mahmoud Daneshmand: A Survey on the Metaverse: The State-of-the-Art, Technologies, Applications, and Chal lenges. IEEE

Internet of Things Journal 10(16), 14671–14688 (2023) https://doi.org/10.1109/jiot.2023.3278329 . Chap. 0. MAG ID: 4377235537 S2ID: db345702ecd677ca216698617a6f74f294262662 — RAYYAN-INCLUSION: "Amir"=¿"Included"

[53] Marchesoni, E.: The current state of metaverse interoperability: Where design framework must go from here. https://venturebeat.com/virtual/the-current-state-of-metaverse-interoperability-where-design-framework-must-go-from-here/ Accessed 2023-02-01

[54] Lee, A.: How Estonia's digital evolution could set it up for the metaverse (2022). https://digiday.com/marketing/how-estonias-digital-evolution-could-set-it-up-for-the-metaverse/ Accessed 2023-02-02

[55] Harkavy, L., Lazzarin, E., Simpson, A.: 7 Essential Ingredients of a Metaverse (2022). https://future.com/7-essential-ingredients-of-a-metaverse/ Accessed 2022-09-02

[56] Giuseppe Macario: WebXR, A-Frame and Networked-Aframe as a Basis for an Open Metaverse: A Conceptual Architecture. arXiv.org (2024) https://doi.org/10.48550/arxiv.2404.05317 . Chap. 0. ARXIV ID: 2404.05317 S2ID: 1e52f7be483e809d18c4e20fde4a06207f62d70f — RAYYAN-INCLUSION: "Amir"=¿"Included"

[57] Bar-Zeev, A.: Beyond Meta: The Seven Verses (2021)

[58] Johnson, S.: Gen Z and the Metaverse (2022). https://www.ipsos.com/en-uk/gen-z-and-metaverse Accessed 2023-01-05

[59] Rea, A.: Identity and the Metaverse (2021). https://medium.com/sylo-io/identity-and-the-metaverse-7e7ac65fc3bd Accessed 2023-01-05

[60] Insights, M.T.R.: Identity protection is key to metaverse innova tion (2022). https://www.technologyreview.com/2022/09/12/1058086/identity-protection-is-key-to-metaverse-innovation/?utm_source=linkedin& utm_medium=tr_social&utm_campaign=Teleperformance E-Brief 2 9.12.22

Accessed 2023-01-05

[61] Intelligence., W.T.: New trend report: Into the Metaverse (2021). https://www.wundermanthompson.com/insight/new-trend-report-into-the-metaverse?j=61174&sfmc_sub=37405083&l=65 HTML&u=4069500&mid=110005021&jb=9008 Accessed 2023-01-05

[62] Ali, M., Naeem, F., Kaddoum, G., Hossain, E.: Metaverse Communications, Networking, Security, and Applications: Research Issues, State-of-the-Art,

and Future Directions. arXiv preprint arXiv:2212.13993 (2022)

[63] Felix, A.: An accessible, disability-inclusive Metaverse? Section: Accessibility (2022). https://www.edf-feph.org/an-accessible-disability-inclusive-metaverse/ Accessed 2023-01-08

[64] Tamuley, P.: Accessibility in the Metaverse: Time to Think (2022). https://www.linkedin.com/pulse/accessibility-metaverse-time-think-priyanka-tamuley/ Accessed 2023-02-01

[65] Hillmann, C.: The Rise of UX and How It Drives XR User Adoption. In: UX for XR: User Experience Design and Strategies for Immersive Technologies, pp. 73–116. Springer, ??? (2021)

[66] Li, C.: Who will govern the metaverse? (2022). https://www.weforum.org/agenda/2022/05/metaverse-governance/ Accessed 2023-02-01

[67] W3C: XR Accessibility User Requirements (2021). https://www.w3.org/TR/xaur/ Accessed 2023-02-01

[68] Zhao, Y., Bennett, C.L., Benko, H., Cutrell, E., Holz, C., Morris, M.R., Sinclair, M.: Enabling people with visual impairments to navigate virtual reality with a haptic and auditory cane simulation. In: Proceedings of the 2018 CHI Conference on Human Factors in Computing Systems, pp. 1–14 (2018)

[69] Jones, D., Ghasemi, S., Graˇcanin, D., Azab, M.: Privacy, safety, and secu rity in extended reality: user experience challenges for neurodiverse users. In: International Conference on Human-Computer Interaction, pp. 511–528 (2023). Springer

[70] Laubheimer, P.: Flexibility and Efficiency of Use: The 7th Usabil ity Heuristic Explained (2020). https://www.nngroup.com/articles/flexibility-efficiency-heuristic/ Accessed 2023-02-01

[71] Sang-Min Park, Sang Min Park, Young-Gab Kim, Young-Gab Kim: A Meta verse: taxonomy, components, applications, and open challenges. IEEE Access, 1–1 (2022) https://doi.org/10.1109/access.2021.3140175 . Chap. 0. MAG ID:

[72] Kreger, A.: UX Case Study: Metaverse Banking VR / AR Design Concept of the Future. 1/15/2023

[73] Reese, L.: UX Design for the Metaverse: Human Factors (2022).

https://www.youtube.com/watch?v=6rjYMKHpl-g&list=PLBVzInsPnP7cfg0pMHq3w8ECQPG0Ta2D&index=2

[74] Nguyen, B.: Internal memos reportedly say Mark Zuckerberg's big metaverse app is suffering a 'quality' problem, and even employees aren't using it enough (2022). https://www.businessinsider.com/mark-zuckerberg-metaverse-app-horizons-quality-problem-report-2022-10?utm_source=linkedin.com&utm_campaign=sf-bi&utm_medium=social Accessed 2023-02-01

[75] Iqbal, H., Latif, S., Yan, Y., Yu, C., Shi, Y.: Reducing arm fatigue in virtual reality by introducing 3D-spatial offset. IEEE Access 9, 64085–64104 (2021). Publisher: IEEE

[76] Milgram, P., Kishino, F.: A taxonomy of mixed reality visual displays. IEICE TRANSACTIONS on Information and Systems 77(12), 1321–1329 (1994). Publisher: The Institute of Electronics, Information and Communication Engineers

[77] O'Hagan, J., Williamson, J.R.: Reality aware vr headsets. In: Proceedings of the 9TH ACM International Symposium on Pervasive Displays, pp. 9–17 (2020)

[78] Delouya, S.: I tried a new interactive 'metaverse' app during a Los Angeles Rams football game. It was cool, but the excitement on the field was more entertaining than the virtual world. (2022). https://www.businessinsider.com/los-angeles-rams-metaverse-at-live-nfl-football-game-photos-2022-12?utm_medium=social&utm_source=linkedin.com&utm_campaign=sf-bi#beatty-said-he-plans-to-add-a-feature-where-app-users-can-play-catch-with-strangers-across-the-stad Accessed 2023-01-05

[79] BRADSELL, L.: UX/UI AND THE METAVERSE (2023). https://www.savantrecruitment.com/insights/ux-ui-and-the-metaverse Accessed 2023-02-10

[80] Lee, H.J., Gu, H.H.: Empirical Research on the Metaverse User Experience of Digital Natives. Sustainability 14(22), 14747 (2022). Publisher: MDPI

[81] Parise, D.: (2022). https://www.linkedin.com/posts/daniloparise-metaverse-ux-uxui-activity-6977957390616780800-5OI1/?utm_source=share&utm_medium=member_desktop Accessed 2023-02-01

[82] Hackl, C.: Metaverse Commerce: Understanding The New Virtual To Physical And Physical To Virtual Commerce Models (2022). https://www.forbes.com/sites/cathyhackl/2022/07/05/

metaverse-commerce-understanding-the-new-virtual-to-physical-and-physical-to-virtual-commerce-models/ ?sh=1331f5255632 Accessed 2023-01-05

[83] Cline, S.: How UX Design Can Help Brands Thrive in the Metaverse (2023)

[84] Zhang, L., Bowman, D.A., Jones, C.N.: Enabling immunology learning in virtual reality through storytelling and interactivity. In: Virtual, Augmented and Mixed Reality. Applications and Case Studies: 11th International Conference, VAMR 2019, Held as Part of the 21st HCI International Conference, HCII 2019, Orlando, FL, USA, July 26–31, 2019, Proceedings, Part II 21, pp. 410–425. Springer, ??? (2019)

[85] Davari, S., Lu, F., Li, Y., Zhang, L., Lisle, L., Feng, X., Blustein, L., Bowman, D.A.: Integrating Everyday Proxy Objects in Multi-Sensory Virtual Reality Storytelling (2021)

[86] Azuma, R.: 11 Location-Based mixed and augmented reality storytelling. Citeseer (2015)

[87] Kit, K.T.: Sustainable engineering paradigm shift in digital architecture, engineering and construction ecology within metaverse. International Journal of Computer and Information Engineering 16(4), 112–115 (2022)

[88] Gong, X., Ren, J., Wang, X., Zeng, L.: Technical Trends and Competitive Situation in Respect of Metahuman—From Product Modules and Technical Topics to Patent Data. Sustainability 15(1), 101 (2022). Publisher: MDPI

[89] Tan-Gillies, H.: L'Or´eal Travel Retail introduces Virtual KOLs and launches 'phygital' campaigns from Viktor&Rolf and Lancˆome (2023). https://www.moodiedavittreport.com/loreal-travel-retail-introduces-virtual-kols-and-launches-phygital-campaigns-from-viktorrolf-and-lancome/ Accessed 2023-02-01

[90] Caballero, I.: Shaping the Metaverse (2022). https://equinoxailab.ai/shaping-the-metaverse/ Accessed 2023-01-05

[91] Tsang, M., Fitzmaurice, G., Kurtenbach, G., Khan, A.: Game-like navigation and responsiveness in non-game applications. Communications of the ACM 46(7), 56–61 (2003). Publisher: ACM New York, NY, USA

[92] Lehtonen, M.: Simulations as mental tools for network-based group learning. In: E-Training Practices for Professional Organizations: IFIP TC3/WG3. 3 Fifth Working Conference on eTRAIN Practices for Professional Organizations (eTrain 2003) July 7–11, 2003, Pori, Finland, pp. 11–17. Springer, ??? (2005)

[93] Roussou, M.: Learning by doing and learning through play: an exploration of interactivity in virtual environments for children. Computers in Entertainment

(CIE) 2(1), 10–10 (2004). Publisher: ACM New York, NY, USA

[94] Mourtzis, D., Panopoulos, N., Angelopoulos, J., Wang, B., Wang, L.: Human centric platforms for personalized value creation in metaverse. Journal of Manufacturing Systems 65, 653–659 (2022). Publisher: Elsevier

[95] Krishnamurthy, R., Chawla, V., Venkatramani, A., Jayan, G.: Transform ing Your Brand Using the Metaverse: Eight Strategic Elements to Plan For. California Review Management (2022)

[96] Steuer, J., Biocca, F., Levy, M.R., *et al.*: Defining virtual reality: Dimensions determining telepresence. Communication in the age of virtual reality 33, 37–39 (1995)

[97] White-Gomez, A.: Inside Water and Music Season 2: Music and the Metaverse (2022). https://www.one37pm.com/nft/water-and-music-season-2 Accessed 2023-01-05

[98] Slater, M., Usoh, M., Steed, A.: Depth of presence in virtual environments. Pres ence: Teleoperators & Virtual Environments 3(2), 130–144 (1994). Publisher: MIT Press One Rogers Street, Cambridge, MA 02142-1209, USA journals-info . . .

[99] Regenbrecht, H., Schubert, T.: Measuring presence in augmented reality environ ments: design and a first test of a questionnaire. arXiv preprint arXiv:2103.02831 (2021)

[100] Thomas, N.J., Baral, R., Crocco, O.S., Mohanan, S.: A framework for gam ification in the metaverse era: How designers envision gameful experience. Technological Forecasting and Social Change 193, 122544 (2023)

[101] Seidel, S., Yepes, G., Berente, N., Nickerson, J.V.: Designing the metaverse. In: Proceedings of the 55th Hawaii International Conference on System Sciences (2022)

[102] Lesmes, L., Hellberg, F.: A Manifesto for the Metaverse (2022). https://metropolismag.com/viewpoints/a-manifesto-for-the-metaverse/ Accessed 2023- 01-05

[103] Weinberger, M., Gross, D.: A Metaverse Maturity Model. Global Journal of Computer Science and Technology 22(H2), 39–45 (2022)

[104] Weinberger, M.: What Is Metaverse?—A Definition Based on Qualitative Meta Synthesis. Future Internet 14(11), 310 (2022). Publisher: MDPI

[105] Zefeng Chen, Wensheng Gan, Jing Sun, Jiayang Wu, Philip S. Yu: Open

Meta verse: Issues, Evolution, and Future. The Web Conference (2023) https://doi. org/10.48550/arxiv.2304.13931 . Chap. 0. ARXIV ID: 2304.13931 MAG ID: 4367393530 S2ID: c53cb927065fac6ffb1c2214722ec99ba766e4dc — RAYYAN INCLUSION: "Amir"=¿"Included"

[106] Lutz, S.: Digitalax Is Racing to Build an OS for Digi tal Fashion in the Metaverse (2021). https://decrypt.co/84141/digitalax-is-racing-to-build-an-os-for-digital-fashion-in-the-metaverse

[107] Guan, J., Irizawa, J., Morris, A.: Extended reality and internet of things for hyper-connected metaverse environments. In: 2022 IEEE Conference on Virtual Reality and 3D User Interfaces Abstracts And Workshops (VRW), pp. 163–168. IEEE, ??? (2022)

[108] Dent, S.: Sony steps into the Metaverse with the 'Mocopi' motion tracking system (2022). https://www.engadget.com/sony-mocopi-movement-tracker-metaverse-avatars-131721036.html?guccounter=1 Accessed 2023-01-05

[109] Park, S., Kim, S.P., Whang, M.: Individual's social perception of virtual avatars embodied with their habitual facial expressions and facial appearance. Sensors 21(17), 5986 (2021). Publisher: MDPI

[110] Siniarski, B., De Alwis, C., Yenduri, G., Huynh-The, T., GUr, G., Gadekallu, T.R., Liyanage, M.: Need of 6G for the Metaverse Realization. arXiv preprint arXiv:2301.03386 (2022)

[111] Nokia's vision of the future is a world where the meta verse replaces smartphones (2022). https://www.rtnewstoday.com/nokias-vision-of-the-future-is-a-world-where-the-metaverse-replaces-smartphones/ Accessed 2023-01-05

[112] Saxena, A.: Walking on water and through pillars: Metaverse app Spatial fails to impress (2022). https://yourstory.com/2022/08/metaverse-app-spatial-people-interact-art Accessed 2023-01-05

[113] Moniruzzaman, M., Chowdhury, F., Ferdous, M.S.: Examining usability issues in blockchain-based cryptocurrency wallets. In: Cyber Security and Computer Sci ence: Second EAI International Conference, ICONCS 2020, Dhaka, Bangladesh, February 15-16, 2020, Proceedings 2, pp. 631–643. Springer, ??? (2020)

[114] Helal, A.E., Marco Costa, T.: Branding in the Metaverse. PhD thesis, Lund University (2022)

[115] Huang, Y., Li, Y.J., Cai, Z.: Security and privacy in metaverse: A comprehensive survey. Big Data Mining and Analytics 6(2), 234–247 (2023)

[116] Lastowka, G.: User-generated content and virtual worlds. Vand. J. Ent. & Tech. L. 10, 893 (2007). Publisher: HeinOnline

[117] Abdelghafar, S., Ezzat, D., Darwish, A., Hassanien, A.E.: Metaverse for brain computer interface: towards new and improved applications. In: The Future of Metaverse in the Virtual Era and Physical World, pp. 43–58. Springer, ??? (2023)

[118] McKinsey and Company: Value Creation in the metaverse (2022). https://www. mckinsey.com/ Accessed 2023-01-05

[119] Sun, J., Gan, W., Chao, H.-C., Yu, P.S.: Metaverse: Survey, applications, security, and opportunities. arXiv preprint arXiv:2210.07990 (2022)

[120] Dwivedi, Y.K., Hughes, L., Baabdullah, A.M., Ribeiro-Navarrete, S., Giannakis, M., Al-Debei, M.M., Dennehy, D., Metri, B., Buhalis, D., Cheung, C.M., *et al.*: Metaverse beyond the hype: Multidisciplinary perspectives on emerging chal lenges, opportunities, and agenda for research, practice and policy. International Journal of Information Management 66, 102542 (2022). Publisher: Elsevier

[121] Zhang, D., Chadwick, S., Liu, L.: The Metaverse: Opportunities and Challenges for Marketing in Web3. Available at SSRN 4278498 (2022)

[122] Dwivedi, Y.K., Hughes, L., Wang, Y., Alalwan, A.A., Ahn, S.J., Balakrishnan, J., Barta, S., Belk, R., Buhalis, D., Dutot, V., et al.: Metaverse marketing: How the metaverse will shape the future of consumer research and practice. Psychology & Marketing (2022). Publisher: Wiley Online Library

[123] Joy, A., Zhu, Y., Pe˜na, C., Brouard, M.: Digital future of luxury brands: Meta verse, digital fashion, and non-fungible tokens. Strategic change 31(3), 337–343 (2022). Publisher: Wiley Online Library

[124] Weking, J., Desouza, K.C., Fielt, E., Kowalkiewicz, M.: Metaverse-enabled entrepreneurship. Journal of Business Venturing Insights 19, 00375 (2023). Publisher: Elsevier

[125] Nidhi Phutela, Priya Grover: Metaverse as a Platform for Student Engage ment in MOOCs (Massive Open Online Courses). International Con ference on Communication and Electronics Systems (2023) https://doi. org/10.1109/icces57224.2023.10192749 . Chap. 0. MAG ID: 4385444417 S2ID: aace01f02cee0d0ae07ea33e141672d7bcec3971 — RAYYAN-INCLUSION: "Amir"=¿"Included"

[126] Yawised, K., Apasrawirote, D., Boonparn, C.: From traditional business shifted towards transformation: The emerging business opportunities and challenges

in 'Metaverse'era. INCBAA 162, 175 (2022)

[127] Festa, G., Melanthiou, Y., Meriano, P.: Engineering the Metaverse for Innovat ing the Electronic Business: A Socio-technological Perspective. In: Thrassou, A., Vrontis, D., Efthymiou, L., Weber, Y., Shams, S.M.R., Tsoukatos, E. (eds.) Business Advancement Through Technology Volume II: The Changing Land scape of Industry and Employment, pp. 65–86. Springer, Cham (2022). https: //doi.org/10.1007/978-3-031-07765-4 4

[128] Aloqaily, M., Bouachir, O., Karray, F., Al Ridhawi, I., El Saddik, A.: Integrating digital twin and advanced intelligent technologies to realize the metaverse. IEEE Consumer Electronics Magazine 12(6), 47–55 (2022)

[129] Koulamas, C., Kalogeras, A.: Cyber-physical systems and digital twins in the industrial internet of things [cyber-physical systems]. Computer 51(11), 95–98 (2018). Publisher: IEEE

[130] Yijing Lin, Hongyang Du, Dusit Niyato, Jiangtian Nie, Jiayi Zhang, Yanyu Cheng, Zhaohui Yang: Blockchain-Aided Secure Semantic Com munication for AI-Generated Content in Metaverse. IEEE Open Jour nal of the Computer Society 4, 72–83 (2023) https://doi.org/10.1109/ojcs. 2023.3260732 . Chap. 0. ARXIV ID: 2301.11289 MAG ID: 4360770953 S2ID: 725303be9d8fc42bb9f8366cb5ec64915e45d17c — RAYYAN-INCLUSION: "Amir"=¿"Included"

[131] Ghiur˜au, T.: How Unity + Volvo Cars built a 3D Test Track (2022). https://medium.com/volvo-cars-engineering/ how-unity-volvo-cars-built-a-3d-test-track-d41fb10bcbf0 Accessed 2023-01-05

[132] Paul, J.: Who Is Kyra? India's First 21-Year-Old Virtual Influencer Who Is Taking The Internet By Storm (2022). https://in.mashable.com/tech/32914/ who-is-kyra-indias-first-21-year-old-virtual-influencer-who-is-taking-the-internet-by-storm Accessed 2023-01-05

[133] SEN GUPTA, M.: The biggest metaverse influencers you should know about (2022). https://www.lifestyleasia.com/bk/tech/influencers-in-the-metaverse/ Accessed 2023-01-05

[134] Deoghuria, S.: 10 Meta Influencers – What Are They And Are AI Influencers On Instagram The Future? (2023). https://blinkstore.in/blog/ meta-influencers-and-ai-influencers/

[135] Hollensen, S., Kotler, P., Opresnik, M.O.: Metaverse–the new marketing uni verse. Journal of Business Strategy (ahead-of-print) (2022). Publisher: Emerald Publishing Limited

[136] Ernest, M.: Gucci partners with roblox to launch 'guccitown' metaverse world (2022). https://www.inverse.com/input/style/gucci-roblox-metaverse-world-gucci-town Accessed 2023-01-05

[137] Repko, M.: RETAIL Walmart enters the metaverse with Roblox experi ences aimed at younger shoppers (2022). https://www.cnbc.com/2022/09/26/walmart-enters-the-metaverse-with-roblox.html Accessed 2023-01-05

[138] Hirsch, P.B.: Adventures in the metaverse. Journal of Business Strategy (ahead of-print) (2022). Publisher: Emerald Publishing Limited

[139] Pymnts: Retailers Try to Spiff up Metaverse Shopping Experi ence (2022). https://www.pymnts.com/artificial-intelligence-2/2022/retailers-try-to-spiff-up-metaverse-shopping-experience/ Accessed 2023-01-05

[140] Uzialko, A.: 8 Ways to Put Pokemon Go to Work for Your Small Business (2022). https://www.businessnewsdaily.com/9228-pokemon-go-business-strategy.html Accessed 2023-01-05

[141] Filipchuk, Y.: [LinkedIn post about Meta verse] (2022). https://www.linkedin.com/posts/yfilip metaverse-competitiveedge-virtualevents-activity-7011746059026640896-35Ve/?utm_source=share&utm_medium=member_desktop Accessed 2023-01-05

[142] Verghese, K.: Travels with my mouse (n.d.). http://www.smarttravelasia.com/secondlife.htm Accessed 2023-01-05

[143] S. Monaco, Giovanna Sacchi: Travelling the Metaverse: Potential Benefits and Main Challenges for Tourism Sectors and Research Applications. Sustainabil ity 15(4), 3348–3348 (2023) https://doi.org/10.3390/su15043348 . Chap. 0. MAG ID: 4320496625 S2ID: d7a84369d5ec6c7096a495a81008966b3bd76c21 — RAYYAN-INCLUSION: "Amir"=¿"Included"

[144] Greener, R.: Accenture Orders Record 60,000 Oculus Headsets (2021). https://www.xrtoday.com/virtual-reality/accenture-orders-record-60000-oculus-headsets/ Accessed 2023-01-05

[145] Schaeffler, M.W., O'Connor, A.: So, You're Looking for a Chief Meta verse Officer (2022). https://www.spencerstuart.com/research-and-insight/so-youre-looking-for-a-chief-metaverse-officer Accessed 2023-01-05

[146] Group., A.: 11 Metaverse Jobs That Will Exist by 2030 | Metaverse and the World of Work (2022). https://www.adeccogroup.com/future-of-work/

latest-insights/11-metaverse-jobs-that-will-exist-by-2030/ Accessed 2023-01-05

[147] Saha, W.: 18 Metaverse Jobs That Could be Hot in the Future (2023). https://geekflare.com/metaverse-jobs/ Accessed 2023-01-05

[148] Breene, K.: What is 'networked readiness' and why does it matter? (2016). https://www.weforum.org/agenda/2016/07/what-is-networked-readiness-and-why-does-it-matter/

[149] Cann, O.: What is competitiveness? (2016). https://www.weforum.org/agenda/2016/09/what-is-competitiveness/ Accessed 2023-01-05

[150] Robertson, D.: The arms race to build the metaverse (2022). https://www.politico.com/newsletters/digital-future-daily/2022/05/04/the-arms-race-to-build-the-metaverse-00030029 Accessed 2023-01-05

[151] Hui, M.: China is eyeing the metaverse as the next internet battleground (2021). https://qz.com/2089316/china-sees-the-metaverse-as-the-next-internet-battleground Accessed 2023-01-05

[152] Kshetri, N.: National Metaverse Strategies. Computer 56(2), 137–142 (2023). Publisher: IEEE

[153] Team, S.: Seoul launches first phase of its metaverse (2023). https://www.smartcitiesworld.net/citizen-engagement/seoul-launches-first-phase-of-its-metaverse Accessed 2023-02-05

[154] Pandey, K.: The U.S. Military Is Rolling Out Its Own Metaverse (2022). https://www.jumpstartmag.com/the-u-s-military-is-rolling-out-its-own-metaverse/ Accessed 2023-01-05

[155] Wyss, J.: Barbados Is Opening a Diplomatic Embassy in the Meta verse (2021). https://www.bloomberg.com/news/articles/2021-12-14/barbados-tries-digital-diplomacy-with-planned-metaverse-embassy#xj4y7vzkg

[156] Park, D.: South Korea's industrial city to reinvent itself in the metaverse (2022). https://forkast.news/south-koreas-industrial-city-to-reinvent-itself-in-the-metaverse/

[157] Service, E.|.E.P.R.: Metaverse Opportunities, risks and policy implications (2022). https://www.europarl.europa.eu/RegData/etudes/BRIE/2022/733557/EPRS_BRI(2022)733557_EN.pdf

[158] Dick, E.: Public Policy for the Metaverse: Key Takeaways from the 2021 AR/VR Policy Conference (2021). https://itif.org/publications/2021/11/15/

public-policy-metaverse-key-takeaways-2021-arvr-policy-conference/
Accessed 2023-01-02

[159] Jing Yuan, Yongquan Liu, Xiaozhe Han, Aiping Li, Liling Zhao: Educational metaverse: an exploration and practice of VR wisdom teaching model in Chinese Open University English course. Interactive Technology and Smart Educa tion 20(3), 403–421 (2023) https://doi.org/10.1108/itse-10-2022-0140 . Chap. 0. MAG ID: 4377027729 S2ID: 185bc6addd5f77368d3effee8f850410962f693a — RAYYAN-INCLUSION: "Amir"=¿"Included"

[160] Rowley, P.: The Biggest Idea You've Never Heard Of: Virtual Nations (2021). https://www.linkedin.com/pulse/biggest-idea-youve-never-heard-virtual-nations-phil-rowley/ Accessed 2023-12-15

[161] Biasi, M., Murgo, M.: The virtual space of the metaverse and the fiddly identi fication of the applicable labor law. Italian Labour Law e-Journal 16(1), 1–11 (2023)

[162] Schickler, J.: Does the Metaverse Need a Free Trade Agree ment? (2022). https://www.coindesk.com/policy/2022/04/01/does-the-metaverse-need-a-free-trade-agreement/ Accessed 2022-02-05

[163] Suffia, G.: Legal issues of the digital twin cities in the current and upcoming European legislation: Can digital twin cities be used to respond to urbanisation problems? In: Proceedings of the 15th International Conference on Theory and Practice of Electronic Governance, pp. 534–537 (2022)

[164] Narin, N.G.: A content analysis of the metaverse articles. Journal of Metaverse 1(1), 17–24 (2021)

[165] Latif U. Khan, Mohsen Guizani, Dusit Niyato, Ala Al-Fuqaha, M´erouane Debbah: Metaverse for Wireless Systems: Architecture, Advances, Standard ization, and Open Challenges. Internet of Things (2023) https://doi.org/10.48550/arxiv.2301.11441 . Chap. 0. ARXIV ID: 2301.11441 MAG ID: 4318620696 S2ID: e729dca0cc66b503c93e91c3be0748b59259484b — RAYYAN-INCLUSION: "Amir"=¿"Included"

[166] Greenwald, M.: Harnessing the Metaverse: States of All Sizes (2022). https://www.wilsoncenter.org/article/harnessing-metaverse-states-all-sizes Accessed 2023-01-05

[167] Larry Zhou, Jordan Lambert, Yanyan Zheng, Long Zheng, Alec Yen, S. Liu, Vivian Ye, Michael Zhou, David Mahar, John Gibbons, Michael Satterlee: Distributed Scalable Edge Computing Infrastructure for Open Metaverse.

2023 IEEE Cloud Summit (2023) https://doi.org/10.1109/cloudsummit57601.2023.00007 . Chap. 0. MAG ID: 4385730088 S2ID: afc4489c4980f93f798d3493cb2aa96f4855364d — RAYYAN-INCLUSION: "Amir"=¿"Included"

[168] Rosenberg, L.B.: Regulating the Metaverse, a Blueprint for the Future. In: Extended Reality: First International Conference, XR Salento 2022, Lecce, Italy, July 6–8, 2022, Proceedings, Part I, pp. 263–272. Springer, ??? (2022)

[169] Howard, R., Wright, D., Ridder, C.: Investing in Metaverse Real Estate: Mind the Gap Between Recognized and Realized Potential (2022). https://www.gravel2gavel.com/metaverse-real-estate-investing/ Accessed 2023-01-05

[170] Talmon, N., Shapiro, E.: Foundations for grassroots democratic metaverse. arXiv preprint arXiv:2203.04090 (2022)

[171] Schneider, J.W.S.: The Metaverse: Decentralized Autonomous Organizations (DAOs) (2022). https://www.hklaw.com/en/insigJacob%20W.%20S.%20Schneider Accessed 2023-01-15

[172] Balkin, J.M.: How to regulate (and not regulate) social media. J. Free Speech L. 1, 71 (2021). Publisher: HeinOnline

[173] Rosenberg, L.: Regulation of the Metaverse: A Roadmap: The risks and regulatory solutions for largescale consumer platforms. In: Proceedings of the 6th International Conference on Virtual and Augmented Reality Simulations, pp. 21–26 (2022)

[174] Rosati, W.S.G..: Will Section 230 and DMCA Translate to the Metaverse? (2022). https://www.jdsupra.com/legalnews/will-section-230-and-dmca-translate-to-4424066/ Accessed 2023-01-05

[175] Saikali, S.: A Race to the Future: The UAE in the Metaverse (2022). https://agsiw.org/a-race-to-the-future-the-uae-in-the-metaverse/ Accessed 2022-01-05

[176] PASCUAL, M.G.: China's metaverse aims to use high-tech to sup press subversion (2022). https://english.elpais.com/science-tech/2022-09-29/chinas-metaverse-aims-to-use-high-tech-to-suppress-subversion.html Accessed 2023-01-03

[177] Cotriss, D.: The Metaverse Economy: Which Industries and Sectors of the Economy Will the Metaverse Disrupt? (2022). https://www.nasdaq.com/articles/the-metaverse-economy%3A-which-industries-and-sectors-of-the-economy-will-the-metaverse Accessed 2023-01-05

[178] Kuruvilla, T., Eagar, R., Meige, A.: Governing The Metaverse. World Government Summit 2023 in collaboration with Arthur Little (2023)

[179] Kang Zhang, Zhanjian Shao, Yun Lu, Yibin Ying, Wei Sun, Zeyu Wang: Introducing Massive Open Metaverse Course (MOMC) and Its Enabling Technology. IEEE Transactions on Learning Technologies, 1–11 (2023) https://doi.org/10.1109/tlt.2023.3289880 . Chap. 0. MAG ID: 4382203530 S2ID: 387de70816a4ca2cd5949006eaddc63cf8b89df0 — RAYYAN-INCLUSION: "Amir"=¿"Included"

[180] Morris, C.: Citi says metaverse economy could be worth $13 trillion by 2030 (2022). https://fortune.com/2022/04/01/citi-metaverse-economy-13-trillion-2030/ Accessed 2023-01-02

[181] Services, A.F.: Government Enters Metaverse (2022). https://www.accenture.com/content/dam/accenture/final/industry/public-service/document/Accenture-Federal-Technology-Vision-2022-Government-Enters-the-MetaverseNew.pdf Accessed 2023-01-05

[182] Revoredo, T.: Decentralization, DAOs and the current Web3 concerns (2022). https://cointelegraph.com/news/decentralization-daos-and-the-current-web3-concerns Accessed 2023-02-15

[183] Nagaoka, S., Motohashi, K., Goto, A.: Chapter 25 - Patent Statistics as an Innovation Indicator. In: Hall, B.H., Rosenberg, N. (eds.) Handbook of the Economics of Innovation, Volume 2. Handbook of the Economics of Innovation, vol. 2, pp. 1083–1127. North-Holland, ??? (2010). https://doi.org/10.1016/S0169-7218(10)02009-5 . ISSN: 2210-8807. https://www.sciencedirect.com/science/article/pii/S0169721810020095

[184] Daily, I.: Country No 2 in Metaverse Patent Application Count (2023). http://www.iprdaily.com/article/index/17528.html Accessed 2023-03-01

[185] Crowley, A.: The Wealth of Virtual Nations: Videogame Currencies. Springer, ??? (2017)

[186] Klein, J.: Michael Wagner: Building a Virtual Nation-State in the Metaverse (2022). https://www.coindesk.com/business/2022/05/09/michael-wagner-building-a-virtual-nation-state-in-the-metaverse/ Accessed 2023-02-02